\documentclass[final,5p,times,twocolumn]{elsarticle}

\usepackage{amssymb}
\usepackage{amsmath}
\usepackage{graphicx}
\usepackage{multicol}
\usepackage[hyphens]{url}
\usepackage{epstopdf}
\usepackage{epsfig}
\usepackage{caption}
\DeclareCaptionType{copyrightbox}
\usepackage{subcaption}
\usepackage{tabularx}
\usepackage{color}

\usepackage{tikz}
\usepackage{multirow}
\usepackage{pdflscape}
\usepackage{pifont}

\journal{}

\begin{document}

\begin{frontmatter}


\title{DDoS Attacks in Cloud Computing:\\ Issues, Taxonomy, and Future Directions}


\author[GS,MSG]{Gaurav Somani}
\author[MSG]{Manoj Singh Gaur}
\author[DS]{Dheeraj Sanghi} 
\author[MC]{Mauro Conti}
\author[RB]{Rajkumar Buyya}

\address[GS]{Central University of Rajasthan, Ajmer, India}
\address[MSG]{Malaviya National Institute of Technology, Jaipur, India}
\address[DS]{Indian Institute of Technology, Kanpur, India}
\address[MC]{University of Padua, Padua, Italy}
\address[RB]{The University of Melbourne, Melbourne, Australia}

\begin{abstract}
{Security issues related to the cloud computing are relevant to various stakeholders for an informed cloud adoption decision. Apart from data breaches, the cyber security research community is revisiting the attack space for cloud-specific solutions as these issues affect budget, resource management, and service quality. Distributed Denial of Service (DDoS) attack is one such serious attack in the cloud space. In this paper, we present developments related to DDoS attack mitigation solutions in the cloud. In particular, we present a comprehensive survey with a detailed insight into the characterization, prevention, detection, and mitigation mechanisms of these attacks. Additionally, we present a comprehensive solution taxonomy to classify DDoS attack solutions. We also provide a comprehensive discussion on important metrics to evaluate various solutions. This survey concludes that there is a strong requirement of solutions, which are designed keeping utility computing models in mind. Accurate auto-scaling decisions, multi-layer mitigation, and defense using profound resources in the cloud, are some of the key requirements of the desired solutions. In the end, we provide a definite guideline on effective solution building and detailed solution requirements to help the cyber security research community in designing defense mechanisms. To the best of our knowledge, this work is a novel attempt to identify the need of DDoS mitigation solutions involving multi-level information flow and effective resource management during the attack. }
\end{abstract}


\end{frontmatter}

\section{Introduction}
\footnote{Important Information: Please cite this paper as:\\ \textcolor{blue}{Gaurav Somani, Manoj Singh Gaur, Dheeraj Sanghi, Mauro Conti, Rajkumar Buyya, DDoS attacks in cloud computing: Issues, taxonomy, and future directions, Computer Communications, Volume 107, 2017, Pages 30-48, ISSN 0140-3664,  http://dx.doi.org/10.1016/j.comcom.2017.03.010.}}
Cloud computing is a strong contender to traditional IT implementations as it offers low-cost and ``pay-as-you-go'' based access to computing capabilities and services on demand. Governments, as well as industries, migrated their whole or most of the IT infrastructure into the cloud. Infrastructure clouds promise a large number of advantages as compared to on-premise fixed infrastructure. These advantages include on-demand resource availability, pay as you go billing, better hardware utilization, no in-house depreciation losses, and, no maintenance overhead. On the other hand, there is a large number of questions in cloud adopters mind which is discussed in literature~\cite{Cloudsec1}~\cite{Cloudsec2}. Most of these questions are specifically related to data and business logic security~\cite{kaufman}. There are many security related attacks, that are well-addressed for the traditional non-cloud IT infrastructures. Their solutions are now being applied to cloud targeted attacks. As data and business logic is located on a remote cloud server with no transparent control, most security concerns are not similar to their earlier equivalents in non-cloud infrastructures.
\par One of these attacks, which has been a much visible attack is the Denial of Service (DoS) attack~\cite{Addressing}. Traditionally, DoS attackers target the server, which is providing a service to its consumers. Behaving like a legitimate customer, DoS attackers try to flood active server in a manner such that the service becomes unavailable due to a large number of requests pending and overflowing the service queue. A different flavor of DoS is Distributed DoS, or DDoS, where attackers are a group of machines targeting a particular service ~\cite{Mirk}. There is a high rise in the number of reported incidents of DDoS, which makes it one of the most important and fatal threat amongst many~\cite{mansfield2015growth}.  
\par More than 20\% of enterprises in the world saw at least one reported DDoS attack incident on their infrastructure~\cite{kaspersky}. Authors in~\cite{cloudddosnews} show a strong anticipation about the DDoS attackers target shift towards cloud infrastructure and services. Many attacks in last two years support these attack anticipations presented in the report. Amongst many recent attacks, there are few popular attacks which gained a lot of attention in the research community~\cite{cloudddosnews}. Lizard Squad attacked cloud-based gaming services of Microsoft and Sony which took down the services on Christmas day in 2015. Cloud service provider, Rackspace, was targeted by a massive DDoS attack on its services. In an another spectacular attack example, Amazon EC2 cloud servers faced another massive DDoS attack. These attack incidents incurred heavy downtime, business losses and many long-term and short-term effects on business processes of victims.  A report by Verisign iDefense Security Intelligence Services~\cite{2015targetingcloud} shows that the most attacked target of DDoS attacks in the last number of quarters is cloud and SaaS (Software as a Service) sector. 

\noindent More than one-third of all the reported DDoS attack mitigations were on cloud services. {One of the most important consequence of DDoS attack in the cloud is ``economic losses''. Report in~\cite{kaspersky} estimates the average financial loss due to a DDoS attack to around 444K USD. There are other reports by Neustar~\cite{economiclosses}, which presents the economic loss data of Q1, 2015. In this report, the average financial loss is more than 66K USD/hour.} DDoS attacks and their characterization become completely different when applied to the context of the cloud. The difference arises mainly due to the consequences of an attack on the victim server. Infrastructure as a Service (IaaS) clouds run client services inside Virtual Machines (VMs).
\par Virtualization of servers is the key to the elastic and on-demand capabilities of the cloud, where VMs get more and more resources when needed and return unused resources when idle. Cloud computing's heavy adoption trend is due to the on-demand computing and resource availability capabilities. This capability enables the cloud infrastructure to provide profound resources by scaling, as and when there is a requirement on a VM. Therefore, a VM will not experience a resource outage as ample amount of on-demand resources are available in the cloud. This feature of ``elasticity'' or ``auto-scaling'' results into economic losses based DDoS attack which is known as Economic Denial of Sustainability (EDoS) attack or Fraudulent Resource Consumption (FRC) attack~\cite{cohen}. 
\par In this paper, we aim to provide a survey of DDoS attacks in the cloud environment. We also differentiate these attacks with the traditional DDoS attacks and survey various contributions in this space and classify them. For this purpose, we prepare a detailed taxonomy of these works to provide aid to comprehend this survey.

\subsection{Need of a survey on DDoS attack in cloud}
 {There are a number of survey papers available which deal with DDoS attacks, both from the perspective of attacks and mitigation in networks. There are surveys and taxonomies available which include traditional DDoS mitigations methods including attack traceback, attack filtering and attack prevention~\cite{shui}~\cite{PengCSUR}. Taxonomies like~\cite{cloudddosSDN} highlight DDoS in the cloud with the perspective of Software Defined Networks.  Surveys such as~\cite{Osanaiye2016147} focus on the solutions which are designed around traffic and behavior change detection. The following are some of the important requirements for this survey:}
\begin{enumerate}
\item Cloud computing and technologies around it are recent phenomenon. It requires a different treatment regarding the characterization of the attack, detection and prevention. The desirable difference is evident in many recent attack incidents~\cite{cloudddosnews}.
\item There are quite a good number of recent studies available on  DDoS attacks, but there is no specific survey (including surveys on Cloud DDoS attacks) available to consider and gather solutions related to resource management aspects of utility computing. 
\item Economic aspects of the DDoS attack (quoted as EDoS) and its consequences on cloud resource allocation is entirely missing from existing surveys; thus, the solutions specific to these issues are required.
\end{enumerate}

\subsection{Survey Methodology}
\label{surveymethod}
We performed literature collection by doing an exhaustive systematic search on all the indexing databases and collecting a huge number of papers related to the area. An initial scan resulted into a subclass of the collection. Another deep scan resulted in the papers we used in our survey and are used in the taxonomy preparation. We believe that the contributions listed in this survey are exhaustive and lists all the important contributions in the emerging area till date.
 
\subsection{Contributions}
We make following contributions in this paper:
\begin{enumerate}
\item We introduce DDoS attack scenario in infrastructure clouds and identify how various elements of cloud computing are affected by DDoS attacks.
\item We present a detailed survey and taxonomy of solutions of DDoS attacks in cloud computing. Based on the developed taxonomy, we identify weaknesses in the state-of-the-art solution space leading to future research directions. 
\item {For a uniform comparison and verification among attack solutions, we provide a comprehensive set of performance and evaluation metrics.}
\item This paper presents a detailed set of design aspects of effective DDoS mitigation at the end. It includes mitigation strategies at resource allocation level instead of preventive and detection strategies used by existing solutions.  
\item This work would help security researchers to deal with the DDoS differently as compared to the treatment given while considering traditional IT infrastructure.
\end{enumerate}

\begin{figure*}[t]
\begin{center}
\includegraphics [width=0.9\textwidth]{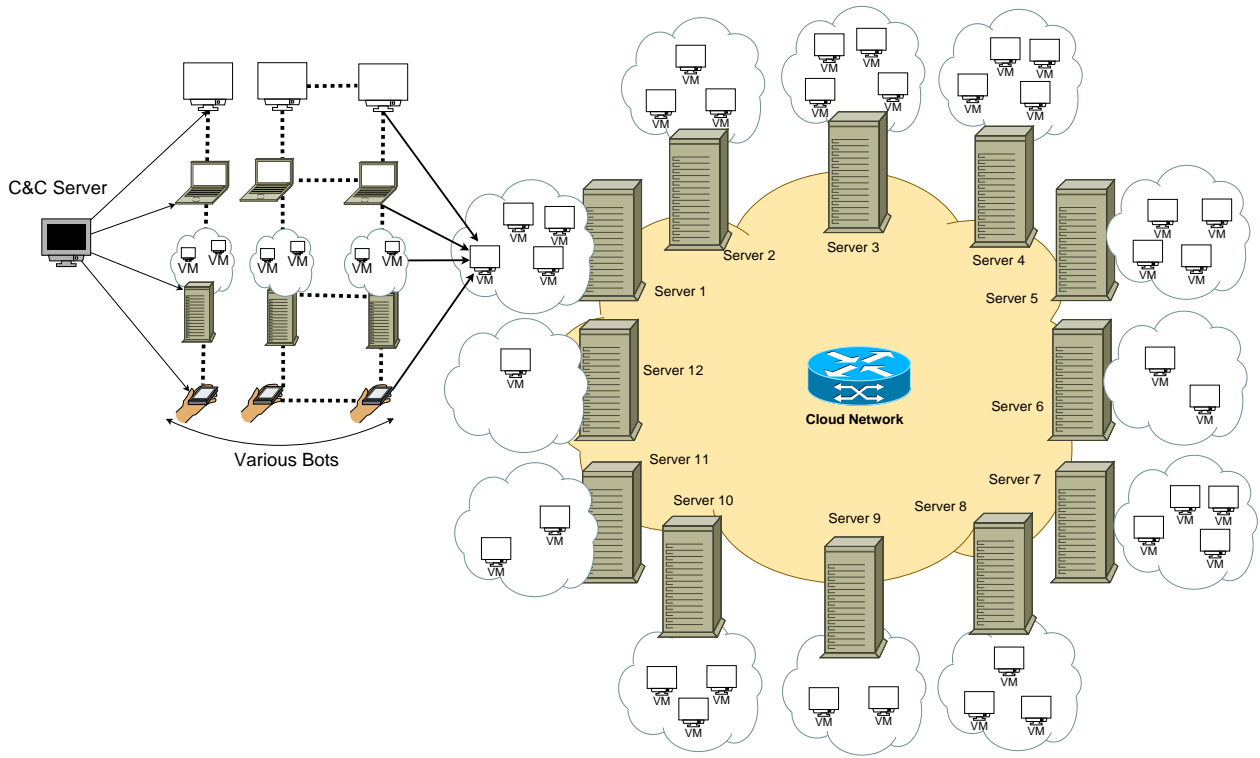}
\caption{DDoS Attack Scenario in Infrastructure Cloud}
\vspace{-2mm}
\label{ddosscenario}
\vspace{-8mm}
\end{center}
\end{figure*}
\subsection{Organization}
We discuss cloud computing and its essential features, which are affected by the DDoS attacks in Section~\ref{Cloudcomputing}. Section~\ref{AttackStatistics} details recent attack statistics to help in understanding the need for this survey. Section~\ref{taxosection} offers a detailed and comprehensive taxonomy to help the reader to understand the broad solution space for DDoS attacks applicable to cloud computing. This taxonomy has three major branches which we discuss in three different sections. These three sections are attack prevention (Section~\ref{prevention}), attack detection (Section~\ref{detection}) and attack mitigation (Section~\ref{mitigation}). In Section~\ref{discussion}, we provide the guideline towards solutions to DDoS attack mitigation. We draw conclusions of this work in Section~\ref{conc}. 

\section{DDoS Attacks and Cloud Computing}
\label{Cloudcomputing}
Cloud computing provides an on-demand utility computing model where resources are available on ``pay-as-you-go'' basis. { In particular, the cloud provider is an ``Infrastructure as a Service (IaaS)'' provider, who provisions VMs on-demand. On the other hand, a service provider is a cloud consumer who has placed the web service in the form of a VM (say an e-commerce application) in the infrastructure cloud provided by the cloud provider.} Figure~\ref{ddosscenario} depicts a typical cloud computing environment with a large number of servers running VMs. 

\subsection{DDoS Attack and Cloud Features}
DDoS attacks have recently been very successful on cloud computing, where the attackers exploit the ``pay-as-you-go'' model~\cite{cloudddosnews}. There are three important features which are the major reasons behind the success trends of cloud computing. On the other hand, the same set of features is proven to be very helpful to DDoS attackers in getting success in the attacks (discussed in Section~\ref{scenario}). We now discuss these three features in detail:
\subsubsection{Auto Scaling} Hardware virtualization provides a feature to shrink-expand resources of a VM while it is running. These properties permit the allocation of additional CPUs, main memory, storage and network bandwidth to a VM when required. Additionally, this can also be used to remove some of the allocated resources when they are idle or not needed. Multiple providers use this resource allocation mechanism with the help of auto scaling~\cite{RAA} web services, which allows cloud consumers to decide the resource need on the basis of resource utilization or similar matrices. The same feature is extended towards adding more VM instances on more physical servers and stopping when there is no need. Machine level scaling (vertical scaling) and data center or cloud level scaling (horizontal scaling) are two crucial features of utility computing. \par Scalability is achieved by spreading an application over multiple physical servers in the cloud. Scalability is driven by high speed interconnects and high speed as well as ample storage. Virtualization of operating systems plays an important role while considering the scalability of VMs. VM cloning and its subsequent deployment are quite fast. Hence, whenever there is a requirement, cloned VMs can be booted on other servers and used to share the load. Scalability is also strongly supported by the live migration of VMs, where a running virtual server can be migrated to another bigger physical server without almost no downtime offering uninterrupted scalable operation.
\subsubsection{Pay-as-you-go accounting} On-demand utility model has become very attractive for consumers due to its leaner resource accounting and billing model. ``Pay-as-you-go'' model allows a cloud consumer to use resources without physically buying them. A VM owner may want to add or remove more resources on-the-fly as and when needed. Other benefits of using cloud platform offer better hardware utilization and no need of arrangements like power, space, cooling and maintenance. Pricing or accounting plays an important role while understanding DDoS attacks in the cloud. Mostly, cloud instances are charged on an hourly basis and thus the minimum accounting period is an hour. Resources can be allotted on fixed basis, pay-as-you-go basis and by auctions. Similarly, storage and network bandwidth are measured using total size and total data (in and out) transfer. It is very clear that these models are ``pay-as-you-go'' models and are still evolving.
\subsubsection{Multi-tenancy} Multi-tenancy gives the benefit of running more than one VMs from different VM owners on a single physical server. Multi-tenancy is a way to achieve higher hardware utilization and thus higher ROI (Return on Investment). An individual user may want to have more than one VMs running similar or different applications on a single physical server.

\subsection{DDoS Attack Scenario in Cloud}
\label{scenario}
A typical attack scenario is as shown in Figure~\ref{ddosscenario}. An infrastructure cloud will have many servers capable of running VMs in multi-tenant virtualized environments. In addition to aiming at  ``Denial of Service'', attackers might aim to attack economic sustainability aspects of cloud consumers. Discussions on this attack have started right after the inception of cloud computing~\cite{cohen}. There are few other contributions where this attack has been termed as Fraudulent Resource Consumption (FRC) attacks~\cite{Idziorek11}. Attackers thoroughly plant bots and trojans on compromised machines over the Internet and target web services with Distributed Denial of Service attacks. DDoS takes the shape of an EDoS attack when the victim service is hosted in the cloud. Organizations exist (also known as ``Booters''), which provide a network of bots to their consumers to plan DDoS attacks on their rival websites~\cite{booters}. Motives of these attacks range from business competition, political rivalry, extortions to cyber wars among countries. 

The cloud paradigm provides enormous opportunities and benefits to consumers and the same set of features are available and may be useful for DDoS attackers. An attacker who plans a DDoS attack would send enough fake requests to achieve ``Denial of Service''. However, this attack would generate heavy resource utilization on the victim server. ``Auto scaling'' \cite{RAA} would take this ``overload'' situation as feedback and add more CPUs (or other resources) to the active pool of resources of this VM. Once a VM gets deployed, it starts as a ``Normal load VM''. Now, let us assume that the DDoS attack has started and consequently VM gets overloaded (``Overloaded VM''). The overload condition triggers auto-scaling features of cloud resource allocation, and it will choose one of the many strategies available in the literature for VM resource allocation, VM migration, and VM placement~\cite{beloglazov}. Overloaded VM may be given some more resources or migrated to a higher resource capacity server or may be supported by another instance started on another server. If there is no mitigation system in place, this process will keep adding the resources. This situation may last till service provider can pay or cloud service provider consumes all the resources. Finally, it will lead to ``Service Denial''. In turn, this leads to on-demand resource billing, and thus economic losses over and above the planned budget may occur. One trivial solution is to run VMs on fixed or static resource profile where the SLA does not have any provision for additional resources on demand. In this case, the DDoS will directly result in ``Denial of Service'' and all the attractive features of the cloud will be lost.
\begin{figure*}[t]
\begin{center}
\includegraphics [width=0.9\textwidth]{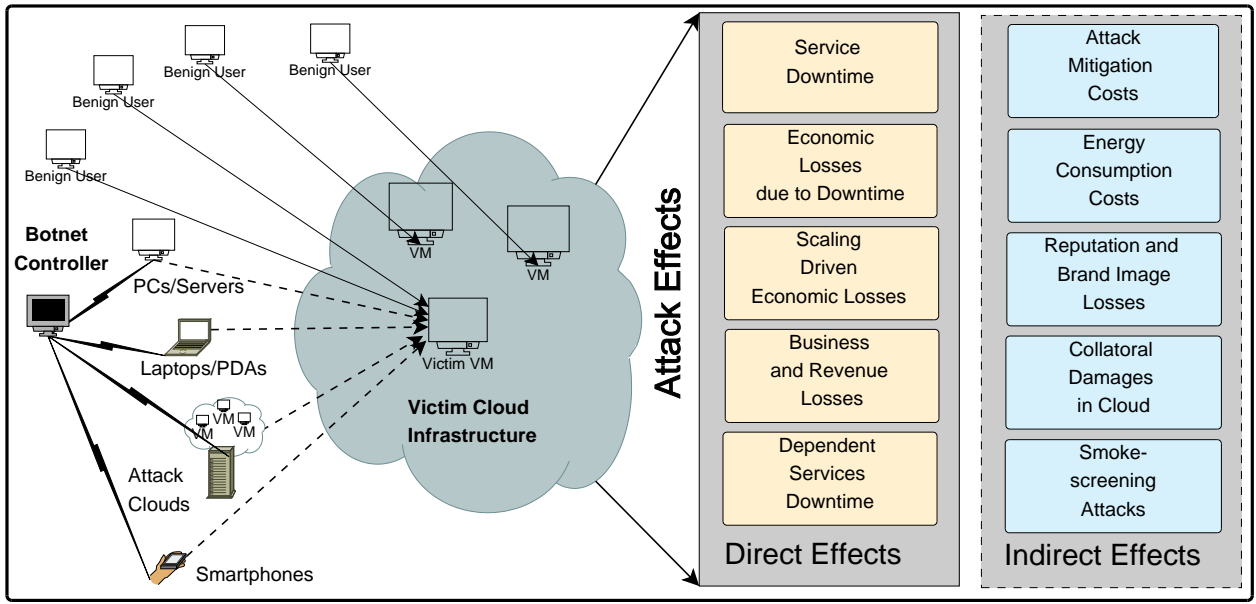}
\vspace{-2mm}
\caption{{DDoS Attack in Cloud: Direct and Indirect Effects}}
\label{effects}
\vspace{-8mm}
\end{center}
\end{figure*}

\section{Attack Statistics and Impact Characterization}
\label{AttackStatistics}
In this section, we provide a coverage of various attack statistics and their impact on various victim organizations. We have also covered few characterization studies to quantify the effects of DDoS attacks in the cloud. {The attack scenario as depicted in Figure~\ref{ddosscenario} can be expanded further in the form of Figure~\ref{effects}. This figure shows details about the attack origin and the resultant attack impacts. The DDoS attacks are mostly botnet driven attacks where a botnet controller directs a large number of automated malware driven bots to launch the attack. We also show cloud originated attacks in the Figure~\ref{effects}. We show directly visible attack effects as well as attack effects which are not directly visible or become visible post-attack. Direct attack effects include service downtime, economic losses due to the downtime, auto-scaling driven resource/economic losses, business and revenue losses, and the downtime and related effects on services which are dependent on the victim service. There are a number of indirect effects to the cloud DDoS attacks. Attack mitigation costs, energy consumption costs, reputation and brand image losses, collateral damages to the cloud components and the effects due to recent smoke-screening attacks. Reputation and brand image losses may not be well quantified and may be treated as long-term losses~\cite{conversion}. Collateral damages include indirect DDoS attacks, addition migrations and scaling, and the energy consumption effects as given in~\cite{somani2016ddos}. We discuss all these attack effects in more detail in this section. }

\subsection{Attack Statistics}
Denial of service attacks are quantified and studied by many security solutions providers in the market~\cite{Akamai}~\cite{neustar}~\cite{prolexic}~\cite{arbor}. There are a number of other reports which state about the impact and rise of DDoS/EDoS attacks in the cloud. It was also anticipated that there will be a major target shift of the DDoS attackers from traditional servers to cloud-based services~\cite{cloudddosnews} and it has even been proven by the Q1 reports of 2015~\cite{2015targetingcloud}. As per this report~\cite{2015targetingcloud}, most of the attack targets were cloud services in Q1, 2015. According to the report by Neustar~\cite{economiclosses}, economic losses per hour at peak times are 470\% more than the previous year. Lizard Squad planned attacks on Microsoft and Sony gaming servers, is the first example. Similarly, Amazon EC2 servers and Rackspace servers, which are cloud service providers, were attacked using a large DDoS attack in early 2015. Economic aspects of these attacks are also challenging. Greatfire.org was targeted by a heavy DDoS attack in March 2015, costing it an enormous bill of \$30,000 daily on Amazon EC2 cloud~\cite{greatfire}. As per report in~\cite{kaspersky}, the average financial damage by a DDoS attack is up to 444,000 USD.  \par Even the innovative ``DDoS as a Service" tools are making it easier for hackers to plan these attacks. As per Q1, 2014 report of ~\cite{prolexic}, the total DDoS attacks within last one year has increased by a significant number (47\%). Another paramount figure to ponder is target servers. More than half of these DDoS attacks targeted towards entertainment and media industry which is mostly hosted in the cloud. {A detailed report regarding the various statistics is covered in~\cite{arbornetworks2015}. As per this report, the DDoS attack bandwidth has grown to more than 500 Gbps in 2016 from just 8 Gbps in 2004.} There are some other reports by Arbor Networks~\cite{arbor}, which state that NTP based reflection and amplification attacks are the new forms of the DDoS. There is an additional attack that is termed very dangerous, has been started showing its effect parallel to a DDoS attack. This attack is known as “Smoke screening” which is an attack to plan information or data breach behind a DDoS. While DDoS distract whole staff in mitigating or preventing from the present situation, the attacker may plan other attacks to harm. 

\noindent As per this report by Neustar~\cite{neustar}, around 50\% of the organizations have suffered from the ``Smoke screening'' attack while they were only mitigating DDoS. Repetition of the attack is also a major issue, and most of the targeted companies (90\%) have faced repetitive attacks leading to vast business losses. The growth and adoption of cloud and DDoS mitigation solutions in the cloud are two major points complementary to each other. Enterprises took few years to start adopting infrastructure clouds after its inception in 2007, and now many of organizations are entirely or partly transformed their IT infrastructure into cloud.

\subsection{DDoS Attack impact studies in the Cloud}
After the inception of the term ``EDoS attack" by Christofer Hoff in 2008~\cite{cohen}, there are some works related to characterization of the DDoS attack in the cloud and study its impacts. To see the effect of the DDoS attack, authors in~\cite{cloudfront} have conducted an important experiment, where they wanted to calculate the maximum possible charges on a cloud service. The authors conducted the experiment by sending 1000 requests/second with 1000 Megabits/second data transfer on a web-service hosted on Amazon CloudFront for 30 days. This experiment accumulated an additional cost of \$42,000 for these additional requests. 

The authors in~\cite{Idziorek11} characterized the effectiveness of the EDoS attack on cloud consumer's bills. Authors in~\cite{Idziorek11} have calculated the additional cost when there is only one attacker that is sending one request per minute for one month. Even this could gather total 13GB of data transfer assuming a normal web request size of 320KB. A similar experiment was conducted in~\cite{vivin} where a web server cluster running on extra-large instance at Amazon EC2 was targeted with an EDoS attack. The observation showed that bills grew on the basis of the number of requests and deployment of additional resources. Authors in~\cite{vlajic2014web} have presented the potential of malicious use of browsers of legitimate users to plan an EDoS attack. 

Authors use social engineering based web-bug enabled spam email to use legitimate browsers for the attack~\cite{vlajic2014web}. Authors have argued that rented bots are easy to detect by the DDoS mitigation infrastructure and web-bugs in the form of a spam email to plan an EDoS attack can be used easily. They planned an attack on Amazon S3 infrastructure and characterized the attack effects. Authors in~\cite{somani2016ddos} have shown the characterization by showing EDoS effects and convergence to DDoS. Similarly, they have conducted a cloud level simulation to show that a DDoS attack in the cloud, may show many side-effects to non-targets including co-hosted VMs, other physical servers, and the whole cloud infrastructure. These effects have formed an important part of the cloud threat and attack model presented in~\cite{bookchapter}. 

\begin{figure*}[thb]
\begin{center}
\includegraphics [width=0.7\textwidth]{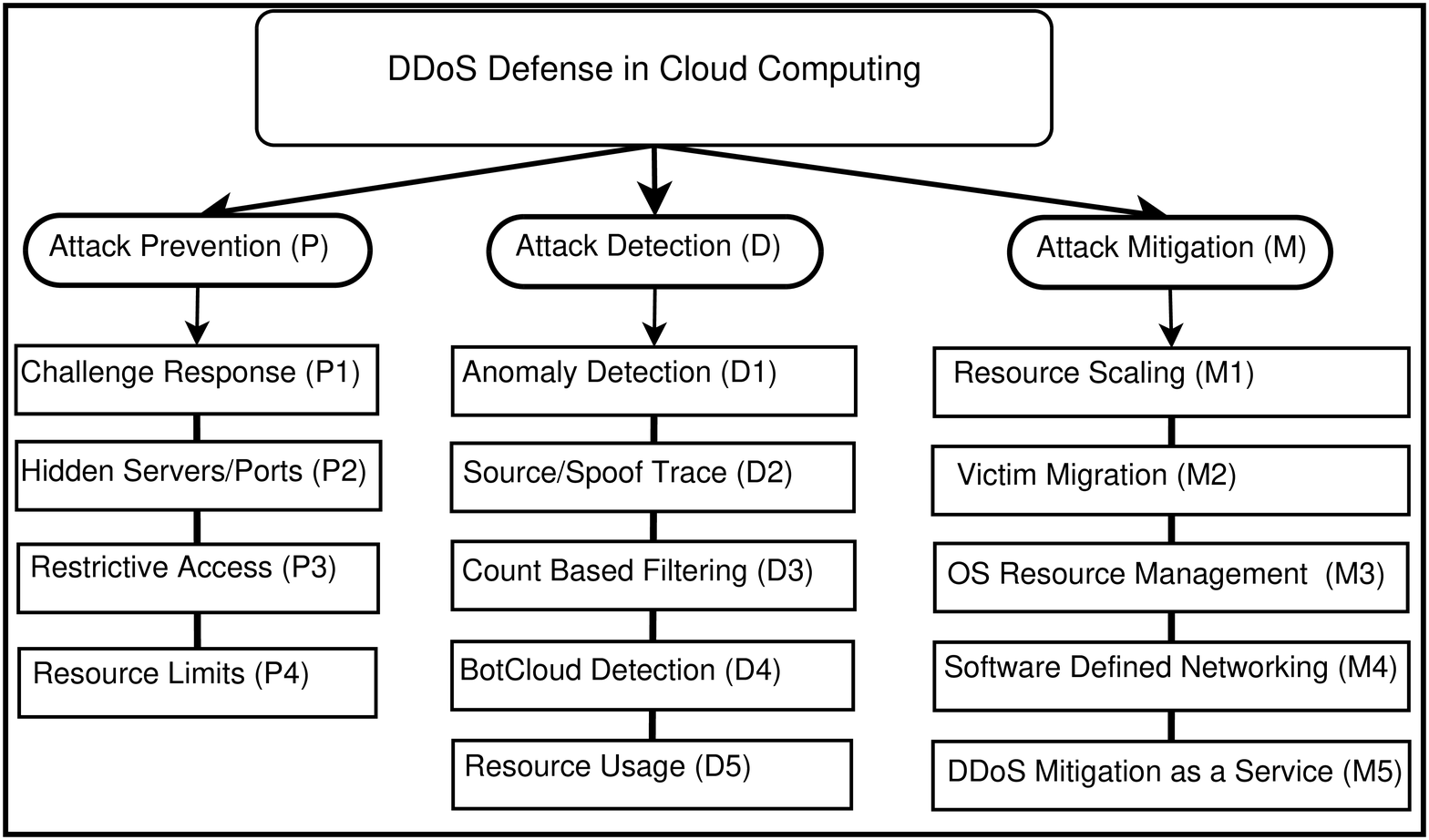}
\vspace{-2mm}
\caption{{DDoS attack prevention, detection and mitigation in cloud: a taxonomy}}
\label{taxo}
\end{center}
\vspace{-6mm}
\end{figure*} 
\begin{figure*}[]
\centering
\includegraphics[width=0.9\textwidth]{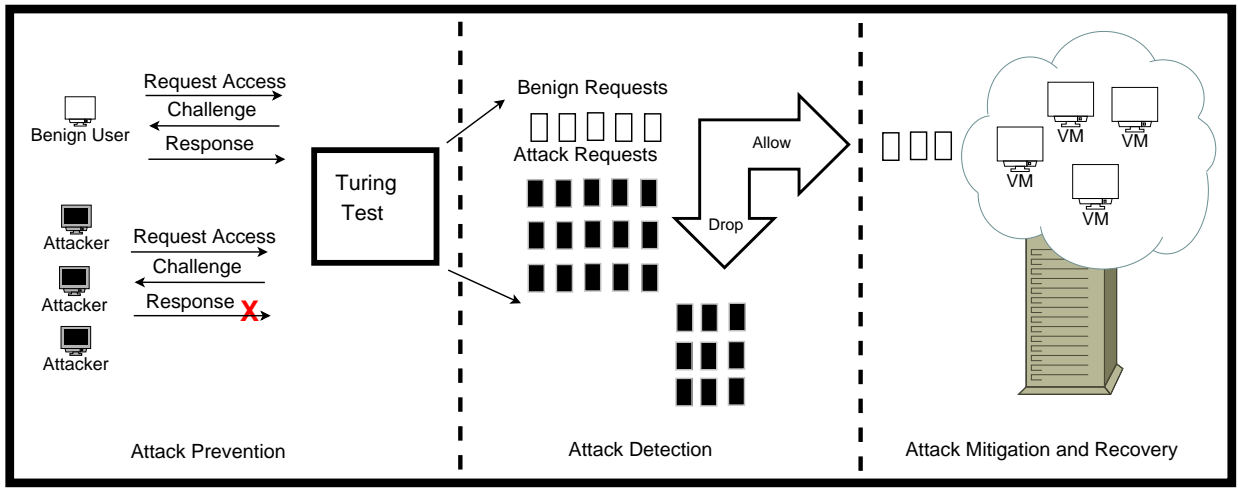}
\vspace{-1mm}
\caption{DDoS Protection in cloud at various levels}
\label{fig:ddosprotection}
\vspace{-3mm}
\end{figure*}

\section{Taxonomy of DDoS Solutions}
\label{taxosection}
This section presents the detailed solution taxonomy of DDoS attacks in the cloud. The final set of contributions in this area were gathered using systematic search methodology discussed in Section~\ref{surveymethod}. The works related to DDoS defense in the cloud have been comprehensively surveyed and prepared as a taxonomy as shown in Figure~\ref{taxo}. To help the particular direction of research, we have included many of works from the DDoS defense in traditional infrastructure. We prepare this taxonomy by keeping a view that this work would serve the purpose of providing a clear, detailed and complete picture of the solutions space, different ideas, and approaches available in the literature. Taxonomy fields are provided a nomenclature to classify different contributions. \par We segment the taxonomy in three important parts which are attack prevention (P), attack detection (D) and attack mitigation and recovery (M). Though, many of these works have contributed in all three or two divisions of this classification, hence, those works are discussed in all those sections individually. The typical solution space looks like the one shown in Figure~\ref{fig:ddosprotection}. 

\noindent At the first instance when the requests come, a simple ``Turing test" may help in preventing the attack. The next stage is anomaly detection to both prevent and detect the attack. There is a large number of contributions in the area of traffic monitoring and analysis. The third stage is based on the methods which are helpful in mitigation as well as recovery. Cloud computing features and profound resources help at this stage. \par We have highlighted the need for more solutions at this stage in section~\ref{mitigation}. There is a large number of contributions available at each stage, and they are listed in the next section. However, in the Figure~\ref{fig:ddosprotection}, we could just show a simplified gist, which misses many other solutions at each stage.  
{Before moving on to the discussion of various DDoS solution categories in the next section, we make an effort to propose important evaluation and performance metrics for various categories of our taxonomy. Table~\ref{metrics} shows the metrics related to the all three subclasses and their subcategories. It is important to highlight that in the next sections, we use these metrics in our discussion to compare the suitability of various solutions. There are many solutions which do not use any evaluation or performance metrics. However, we believe that these important metrics can help the community to orchestrate solutions which are verifiable against the important properties we list in the Table~\ref{metrics}. }

\begin{table*}[t]
\begin{center}
\centering
{
\begin{tabular}{|c|l|l|}\hline
 & {\it \bf Subcategory}&{\it \bf Important metrics to benchmark the solutions}\\ \hline
\multirow{4}{*}{\rotatebox{90}{\parbox{3cm}{ \small \bf Attack \\Prevention}\hspace{-15mm}}} 
& Challenge Response & \small  Accessibility, usability, puzzle generation, storage, and verifiability, and false alerts\\ \cline{2-3}
& Hidden Servers/ports & \small  Redirection, overhead of server replicas and load balancing, and all other puzzle metrics \\ \cline{2-3}
& Restrictive Access & \small  Accessibility, usability, response delay and false positives and negatives in admission control\\ \cline{2-3}
& Resource Limit  & \small Cost and overhead of management of additional reserved resources \\ \hline
\multirow{5}{*}{\rotatebox{90}{\parbox{3cm}{ \small \bf Attack \\Detection}\hspace{-15mm}}} 
& Anomaly Detection & \small Overhead cost of training and profiling and false positives and negatives \\ \cline{2-3}
& Source and Spoof Trace & \small TTL data verification and traceback costs and false positives and negatives   \\ \cline{2-3}
& Count Based Filtering & \small Suitability to various static and dynamic counts in minimizing the false alerts \\ \cline{2-3}
& BotCloud Detection & \small Overhead cost of learning and verifying traffic flows and false alerts \\ \cline{2-3}
& Resource Usage & \small  Overhead of employing monitors and counters, and threshold suitability   \\ \hline
\multirow{5}{*}{\rotatebox{90}{\parbox{3cm}{ \small \bf Attack\\ Mitigation}\hspace{-15mm}}} 
& Resource Scaling & \small Auto-scaling decision and threshold suitability \\ \cline{2-3}
& Victim Migration & \small Migration downtime, costs and network overhead for deltas  \\ \cline{2-3}
& OS Resource Management & \small  Attack mitigation, reporting and downtime, and attack cooling down period \\ \cline{2-3}
& Software Defined Networking & \small Overhead cost of training, profiling, and false positives and negatives  \\ \cline{2-3}
& DDoS Mitigation as a Service & \small Solution costs, service downtime and other metrics based on different solutions \\ \hline
\end{tabular}
}
\end{center}
\vspace{-2mm}
\caption{{Various performance metrics to benchmark the DDoS attack solutions in cloud computing}}
\label{metrics}
\vspace{-3mm}
\end{table*}

\section{Attack Prevention (P)}
\label{prevention}
DDoS prevention in the cloud is a pro-active measure, where suspected attackers' requests are filtered or dropped before these requests start affecting the server. Prevention methods do not have any ``presence of attack'' state as such, which is usually available to the attack detection and mitigation methods. Therefore, prevention methods are applied to all users whether legitimate or illegitimate. Most of these methods are tested against their usability, which incurs an overhead for the server as well as legitimate clients. {We further classify this direction in four subclasses:
\begin{enumerate}
\item Challenge Response.
\item Hidden Servers/ports.
\item Restrictive Access.
\item Resource Limit.
\end{enumerate} 
For a quick view, the overall theme of each set of these methods, their strengths, challenges, and weaknesses are listed in Table~\ref{P1}. We also prepare a list of important individual contributions in Table~\ref{P2}. We enlist a brief theme of each solution to provide an overview about the variety of contributions available in each of the subclass. }

\begin{table*}
\begin{center}
\centering
{
\begin{tabular}{|l|l|l|l|l|}\hline
{\it \bf  Techniques}	&	{\it \bf Strengths }	& 	{\it \bf Challenges} 			& {\it \bf Limitations} & {\it \bf Contributions}\\ \hline
 \parbox{2cm}{Challenge\\ Response (P1)}  		& \parbox{3.5cm}{Effective and usable  methods using puzzles to differentiate human and bots}  & \parbox{3.5cm}{ Overhead of graphics generation and its storage}  &  \parbox{3.5cm}{Image segmentation, OCR, dictionary and parsing attacks, and puzzle accumulation attacks} &  \parbox{2cm}{\cite{spow}\cite{scrubber}\cite{enhanced} \cite{edosshield}\cite{alosaimi}\cite{moving} \cite{comber}\cite{capabilities}}\\\hline

 \parbox{2cm}{Hidden Servers/Ports (P2)}  		& \parbox{3.5cm}{Service is being offered to legitimate users while no direct connection is established with the real server in the first instance}  & \parbox{3.5cm}{Redundant servers ports and load balancing among them is needed}  &  \parbox{3.5cm}{Overhead of additional security layer and redirections} &  \parbox{2cm}{ \cite{moving}\cite{spow}\cite{edosarmor} \cite{jia2014catch}\cite{army} }\\\hline
 
 \parbox{2cm}{Restrictive Access (P3)}  		& \parbox{3.5cm}{ Admission control or instead of blocking/dropping responses are prioritized for different classes of users}  & \parbox{3.5cm}{Quality of service concerns and overhead of maintaining number of connections for  delayed period}  &  \parbox{3.5cm}{Not scalable in case of massive DDoS with spoofing by large number of sources} &  \parbox{2cm}{ \cite{baig}\cite{index}\cite{spow} \cite{edosarmor} }\\\hline
 
\parbox{2cm}{Resource Limits (P4)}  		& \parbox{3.5cm}{ Limiting the economic losses by restricting the maximum usable resources by a VM}  & \parbox{3.5cm}{Determining the resource limits and capacity planning of  a server}  &  \parbox{3.5cm}{It does not prevent DDoS and its effects, except  limiting the economic losses due to cloud billing } &  \parbox{2cm}{\cite{Cloudwatch}\cite{amazondiscussionforum}\cite{awsddosprotection}  }\\\hline
\end{tabular}
}\end{center}
\caption{{DDoS Attack Prevention Techniques in Cloud: P2 Other Prevention Methods}}
\label{P1}
\end{table*}

\subsection{Challenge Response (P1)}
\label{SectionP1}
\par {Challenge-Response Protocols (CRP) are designed to identify the presence of real users. Many times, this concept has been applied in an opposite manner, where the protocol tries to determine if the user is a bot/attacker machine, especially in the case of crypto-puzzles or proof-of-work. One of the most common prevention technique is a Turing test in the form of a CAPTCHA, which is usually one of the most preferred methods in the category of challenge-response protocols. In addition to the methods related to cloud, some important CRPs from traditional DDoS defenses are also added to this discussion. Graphical Turing tests are popular CRP implementations available today. Instead of showing plain text challenge and seeking an answer, these tests may present an image and a question related to that image. The image may have a picture, text with various impurities like an arc, distortion, and noise. Graphical CAPTCHA may have moving images in the form of .GIF or set of multiple pictures to choose from. Crypto puzzles are used to assess the computational capability of a client. Crypto puzzles are questions seeking output of a function with given inputs. For example, let us consider a hash function $f(x,y)$ with inputs $a$ and $b$. The client is expected to compute $f(a,b)$ and return the answer back in some stipulated time.}

Now, we discuss few important strategies related to challenge response schemes to prevent DDoS attack in cloud computing. EDoS Shield~\cite{edosshield} and Alosaimi et al.~\cite{alosaimi} used graphical Turing tests to prevent the bot driven attack from occurring.  Authors in~\cite{alosaimi} proposed a DDoS Mitigation System (DDoS-MS), where initial two packets from the client side, form the basis of the attack identification and subsequent mitigation. In their work, they used both graphical Turing tests and crypto puzzles to identify the attacker. Authors in~\cite{edosshield} proposed a solution that filters requests on the basis of graphical Turing tests (CAPTCHAs). In this mode, a Virtual Firewall (VF) shield is designed which distinguishes the incoming requests on the basis of two lists, white and black. These records are updated on the basis of the success and failures of graphical Turing tests. To prove the effectiveness and novelty of their solution, authors have conducted simulations to show the effect of their scheme on end-to-end delay, cost, and other performance indicators like throughput and bandwidth.  There are a variety of crypto puzzles with different difficulty levels in~\cite{spow}~\cite{scrubber}~\cite{enhanced}. Authors in~\cite{spow} presented sPoW (Self-Verifying Proof of Work) methodology to mitigate EDDoS (Distributed EDoS). They provided a method to mitigate both network-level EDDoS and Application level EDDoS by extending the work proposed in~\cite{capabilities}. In~\cite{capabilities}, instead of accepting all the traffic, they are only accepting the traffic that they are capable of taking. The authors in~\cite{spow} provided an innovative solution where they use crypto-puzzle to identify legitimate customers. These crypto-puzzles are self-verifying and do not run on the server. Instead of the server, the client computes the solution. On the basis of the time taken to solve the crypto-puzzle servers/nodes in the intermediate path, it will be decided whether the incoming traffic is legitimate traffic or not. The salient feature of this approach is that DDoS attacker may send their traffic even at a higher rate by speedily computing the puzzle, even in this case, sPoW approach does not allow the traffic. On the other hand, if DDoS traffic comes at a normal rate (equivalent to the rate at which legitimate customer sends) then their approach is successful in limiting the traffic.

\par {Challenge Response schemes provide an easy way of implementing the attack prevention methods by addressing the most common automated, bot originated and rate based attacks. A list of good qualities crypto puzzles are described in~\cite{tlscrypto}. The crypto puzzle should be solvable in a definite time and should not have other possible methods. Additionally, the server should be able to compute answers and verify them with ease.}

\noindent {Proof-of-work approaches are crypto puzzles but may have advanced features to utilize the client computation power and based on the correctness of solution and time, the authentication, and prioritized access is granted~\cite{spow}~\cite{moving}. This approach has multiple benefits including computation overhead shifting to the client and stopping overwhelming computationally equipped clients. }

\par {Accessibility and conversion rates are two important points, which have been discussed recently against challenge-response protocol implementations specifically, CAPTCHAs~\cite{accessibility}. There are many CRPs, which are designed and tested from the perspective of their attack persistence, accessibility, overhead, puzzle generation, and storage requirements. Many of these are issues related to the area of Human Computer Interaction (HCI). One of the important aspects of ``Challenge-Response Protocols" is {\it Accessibility}, which should be considered while designing the question generation module. Designing difficult questions so that bots cannot construct their answer is quite an easy task, but a normal user should also be able to answer the questions with adequate comfort. Solutions based on Turing tests should be examined using a usability and accessibility study. Text puzzles are known to be cracked using dictionary attacks or parsing attacks. There is a number of limitations which are posed by~\cite{edosshield}, like the puzzle accumulation attack where an attacker sends a large number of requests for getting puzzles but does not solve them. It would result in an extra overhead of generating the puzzles at the server end. These Turing tests require additional overhead to generate graphics and storage space to store images. There are multiple works related to CAPTCHA cracking using image segmentation and optical character recognition (OCR).}

\begin{table}[t]
\begin{center}
\centering
{
\begin{tabular}{|r|c|l|}\hline
\rotatebox{90}{\it \bf \parbox{2cm}{Solution \\category}} & \rotatebox{90}{\it \bf Contribution}&{\it \bf Major theme of the contribution}\\ \hline
\multirow{8}{*}{\rotatebox{90}{\parbox{2.8cm}{ \small \bf Challenge\\Response (P1)} \hspace{-11mm}}} & \cite{spow} & \small Crypto puzzles to identify benign traffic \\ \cline{2-3}
										& \cite{scrubber} & \small Crypto puzzle levels based on the attack rate \\  \cline{2-3}
										&  \cite{enhanced} & \small Crypto puzzles to identify benign  traffic \\  \cline{2-3}
										& \cite{edosshield} &  \small Graphical Turing tests \\  \cline{2-3}
										& \cite{alosaimi} & \small Both graphical as well as crypto puzzles \\  \cline{2-3}
										& \cite{moving} & \small Proof-of-work puzzles \\  \hline
										& \cite{comber} &  \small Turing tests combined with other techniques \\ \cline{2-3}
\multirow{6}{*}{\rotatebox{90}{\parbox{2.8cm}{\small  \bf Hidden Servers/\\Ports (P2)} \hspace{-6mm}}} &\cite{moving} & \small Moving target approach to hide the servers \\ \cline{2-3}
										& \cite{spow} & \small Secure ephemeral servers with authentication\\  \cline{2-3}
										& \cite{edosarmor} & \small Limits number of connections on hidden ports\\  \cline{2-3}
										& \cite{jia2014catch} & \small Moving targets using server replica shuffling \\  \cline{2-3}
										& \cite{army} & \small Hidden server only visible to benign users \\  \cline{2-3}
										& \cite{nakao} & \small Proxy forwards benign requests to the server \\   \hline						
\multirow{4}{*}{\rotatebox{90}{ \parbox{2cm}{\small  \bf Restrictive\\Access (P3)} \hspace{-5mm}}} & \cite{baig} & \small Admission control based on delayed response \\ \cline{2-3}
										& \cite{index} & \small Human behavior (rate) detection and access \\  \cline{2-3}
										& \cite{spow} &  \small Client reputation based prioritized access \\  \cline{2-3}
										& \cite{edosarmor} &  \small  Admission control puzzles and hidden ports \\  \hline
\multirow{3}{*}{\rotatebox{90}{\parbox{2cm}{\small  \bf Resource\\Limits (P4)} \hspace{-5mm}}} &\cite{awsddosprotection}  & \small Resource Scaling to absorb the attack \\ \cline{2-3}
										& \cite{amazondiscussionforum} & \small Resource caps to limit the attack effects \\  \cline{2-3}
										&\cite{Cloudwatch} & \small Cloud metric monitoring and alarms   \\  \cline{2-3}
										&\cite{DARAC} & \small DDoS Aware scaling and capacity planning \\  \hline
\end{tabular}
}
\end{center}
\caption{{DDoS Attack Prevention Techniques in Cloud}}
\label{P2}
\vspace{-5mm}
\end{table}
\subsection{ Hidden Servers/Ports (P2)}
{Hidden servers or hidden resources such as ports is an important method to remove a direct communication link between the client and the server. The objective of hiding the servers, is achieved by keeping an intermediate node/proxy to work as a forwarding authority. The important jobs of this forwarding authority may include balancing the load among the servers, monitoring the incoming traffic for any vulnerability, and fault-tolerance and recovery of the servers. }

Various approaches have differently used the features of hiding the resources, e.g. hidden proxy server \cite{moving}, ephemeral servers \cite{spow} and hidden ports \cite{edosarmor}. Authors in~\cite{moving} proposed a moving target method to defend from DDoS attacks. They proposed the inclusion of many hidden proxy servers which may be dynamically assigned and changed to save legitimate clients. This approach has some practical issues like scalability, the inclusion of large no. of proxy servers, shuffling. Even different web services may not like to have changing server addresses in between connections. This method uses client puzzles using PoW (Proof-of-Work) to distinguish between attackers and normal traffic. Additionally, some of these approaches randomly allocate different hidden servers. Jia et al.~\cite{jia2014catch} have used the moving target based mechanism by shuffling the targets to confuse the attackers. This is achieved using the server replicas. This solution requires the overhead of maintaining the replicas and managing the moving target strategies. Additionally, authors have proposed strategies of effective shuffling assignments of clients requests to servers. Authors in~\cite{army} have proposed a DDoS detection mechanism which is a request rate based detection method. The proposed method black lists incoming client request on the basis of their threshold rate. 

\noindent By this blacklist, access is granted by special ``army nodes" creating a virtual firewall. {Authors have argued that this way, the server could continue to serve legitimate clients. Similarly, authors in~\cite{nakao} have proposed a solution where a proxy server is used to test and forward the benign requests to the server behind the fore-front service. }

{Hidden servers or ports are preventive mechanisms to save the real service to face a DDoS attack. Therefore, requests to these hidden servers or ports are redirected by authentication/proxy servers which is the first server to be encountered by a client. Authentication provides an extra security layer to secure the actual service.  Hidden servers can help in stopping the malicious or massive traffic to affect the real server.  This extra layer may also support other purposes of redirection and load balancing among servers.}

{The major limitation of this approach include the cost of the intermediate servers, time delay and computation overhead of redirection and its management at the intermediate nodes. Additional overhead includes the cost of  maintaining the server replicas and their backup management. }

\subsection{Restrictive Access (P3)}
{Restrictive access techniques are basically admission control methods to take preventive action against the service capacity. Some of these strategies have implemented the prevention by delaying responses/access to the suspected attackers or even additional clients. In many of the contributions, this delay is introduced by prioritizing the legitimate clients or selecting clients with ``good'' past behaviors. There are few techniques which are based on ``Delayed access" and ``Selective access" which are mostly similar, except that the strategies to provide the access to the clients are different.}

In some cases, reputation is the basis of the admission control mechanism, in which some users are preferred over the others on the basis of reputation~\cite{spow}. Reputation is calculated on the basis of the correctness of crypto puzzles within a definite time and past web access behavior. Authors in~\cite{spow} have named it as ``capabilities". Authors in~\cite{baig} gave a solution which did not drop any request based on its behavior, instead, they delayed the access to them. This delayed access prevents the attack to occur and even does not trigger auto-scaling. The proposed method controls the user access requests by their past web access history. In case these claims reach certain thresholds, the request responses are delayed instead of dropping the requests. These thresholds are decided from the request history of users. The effectiveness of delayed responses is questionable in real environments because of user accessibility issues, which requires timely responses. Authors in ~\cite{index} have followed a different approach where if a user does not behave as per typical human behavior, it is blocked for a specific period and then it is again unblocked. Authors have proposed a novel subclass of the DDoS attack and termed it as index page based attack where the very first page or homepage of the website is targeted for a DDoS attack~\cite{index}. Clearly, the first page of every website should be free and can be fetched without solving any puzzle or authentication. Authors have shown attacks on this first page, where no Turing test prevention mechanism may work. Authors have given human behavior based identification to mitigate the attack and drop the requests of an attacker. This way, they serve the attacker, equivalent to no. of times, they serve to legitimate clients. After a certain request count/threshold, they blocked the attacker for some time. There are some contributions which propose to provide access to only those to whom they can provide as per actual resource capability at the server end. 

Instead of queuing all the clients, they~\cite{edosarmor} proposed an admission control algorithm, where a limited number of clients are served simultaneously, who have solved the Turing test and assigned to hidden ports. Once the test is passed by legitimate users, the proposed mechanism tries to limit the number of clients at any time by using an admission control algorithm. This is done by providing service to a limited number of clients on hidden ports using a port key.  The server allocates resources on the basis of priority which is calculated based on the user behavior. The behavior is basically the web behavior on an e-commerce site on the basis of multiple parameters. 

{Most of the admission control methods which implements restrictive access to stop the DDoS attacks to occur are primarily based on delayed access or reputation based access. These methods provide a good way to optimize the server capacity by allowing requests based on the available resources. The ``reputation'' or ``capability'' is calculated based on the past access pattern or the time to solve the crypto puzzles.} 

{On one hand, these input control methods are solely dependent on the server capacity and client capability to compute the puzzle responses.} 

\noindent {At times, this restriction may limit the server to address the accessibility or usability perspective for fresh clients. As discussed in Section~\ref{SectionP1}, the problems associated with the puzzle based solutions are also applicable here. Additionally, in case of sophisticated or stealthy attacks, the malicious attackers may try to earn the ``reputation'' before they show their real malicious behavior. }

\subsection{ Resource Limits (P4)}
{As discussed in Section~\ref{AttackStatistics} on attack characterization, it was visible that the economic bills generated by a DDoS attack can be enormous. Resource limits can help in preventing these economic losses by correct auto-scaling decisions. However, deciding whether the resource surge has come due to the DDoS attack or due to the real genuine traffic, is a very difficult task. Another way to prevent these resource losses is to put fixed resource services or ``capped'' resource limits on each service in the cloud. By doing this, we will miss the advantages of important features of cloud computing such as  on demand resource allocation. }

There are number of discussions and demands by cloud consumers on providing a track of resource utilization in the form of alerts. Additionally, some of the providers, have started providing the real-time monitoring services~\cite{Cloudwatch}. They have also started providing resource limits in the form of ``Caps" on maximum resources a VM would be able to buy and sustain. {There are other solutions such as~\cite{DARAC}, in authors develop a resource allocation algorithm where the resources are only increased if the resource surge is due to the real genuine traffic. }

{It would not help the cloud consumer to stop the DDoS to occur; however, it can surely limit the bills on the cost of service downtime (as the VM would reach the resource limit and would not be able to serve any clients as resource outage will lead to DDoS)~\cite{amazondiscussionforum}.}

{Resource limits can surely restrict the cost penalty on the dynamically scaled resourced but they can also limit the usage of on-demand computing feature of cloud computing. }

{Attack prevention mechanism discussed above present a variety of methods available for the preventive security. However, it is important to note that these prevention mechanisms alone can not help in combating the DDoS attacks in cloud infrastructures. Another line of support from other mechanisms such as detection and mitigation mechanisms is needed once the attack is already present. On the other hand, attack mitigation methods are indeed important first-aid solutions to the overall DDoS solution framework we discuss in Section~\ref{effective}. }

\section{Attack Detection (D)}
\label{detection} 
Attack detection is achieved in a situation where attack signs are present on the server in terms of its services and monitored performance metrics. These attack signs are initial signs, where the attack has just started to take the shape, or there may be a situation, where the attack has already deteriorated the performance. These methods may seem to be similar to ``attack prevention" at times, and many of contributions have provided solutions in the same manner. Various performance metrics, which are monitored and affected due to an attack range from large response times and timeouts to higher memory and CPU usage. {We further classify this section into five subcategories:
\begin{enumerate}[1.]
\item Anomaly Detection.
\item Source and Spoof Trace.
\item Count Based Filtering.
\item BotCloud Detection.
\item Resource Usage.
\end{enumerate} 
For a quick view, the overall theme of each set of the classified methods, their strengths, challenges, and weaknesses are listed in Table~\ref{D1}. We also prepare a list of important individual contributions in Table~\ref{D2} where we enlist a brief theme of each solution to show the variety of contributions available in each subclass.}

\begin{table*}[htb]
\vspace{-3mm}
\begin{center}

\centering
{
\begin{tabular}{|l|l|l|l|l|}\hline
{\it \bf  Techniques}	&	{\it \bf Strengths }	& 	{\it \bf Challenges} 			& {\it \bf Limitations} & {\it \bf Contributions} \\ \hline

 \parbox{2cm}{Anomaly Detection (D1)}  		& \parbox{3.5cm}{Machine learning and feature based detection}  & \parbox{3.5cm}{ Feature identification, testing and minimizing false alarms and IP spoofing }  &  \parbox{4cm}{Scalability issues and overhead of training, matching and statistical analysis of traffic features} &  \parbox{2cm}{\cite{idziorekdetecting}\cite{covariance}\cite{statistical}  \cite{Idziorek11}\cite{shamsolmoali2014statistical}\cite{gomez2000analysis} \cite{templeton2003detecting}\cite{cbf}\cite{jeyanthi}  \cite{edosarmor}\cite{ddosdefense} \cite{timespent}}\\ \hline

\parbox{2cm}{Source and Spoof Trace (D2) }  		& \parbox{3.5cm}{Identifying the source of of web requests to stop spoofing}  & \parbox{3.5cm}{Filtering at edge routers and suitability of TTL based methods}  &  \parbox{4cm}{Cooperative mechanisms require network devices and service support } &  \parbox{2cm}{\cite{yu2016feasible}\cite{chonka}\cite{botnetsurveysilva} \cite{yang}\cite{osanaiye2015short}\cite{sourceauthentication} \cite{multistage}\cite{luo2013preventing}\cite{law2005you}}\\\hline

\parbox{2cm}{Count Based Filtering (D3)}  		& \parbox{3.5cm}{Hop count, number of connections or number of requests based threshold filtering }  & \parbox{3.5cm}{Requires TTL hop data of real user. Heterogeneous implementations of hop count. Deciding on count threshold is  a challenge}  &  \parbox{4cm}{ IP spoofing issues may defeat  the (non-TTL) schemes. Only successful in case of two different  TTLs for same source IPs are received. False alarms. Probing is also needed. } &  \parbox{2cm}{\cite{comber}\cite{enhanced}\cite{edosshield}  \cite{index}\cite{army}\cite{deflate}}\\\hline

\parbox{2cm}{BotCloud Detection (D4)}  		& \parbox{3.5cm}{Detecting the attack sources inside the cloud by monitoring the features of VMs and the network}  & \parbox{3.5cm}{Identifying the activities and their thresholds for various suspicious activities}  &  \parbox{4cm}{Very difficult to detect all kinds of attack flows (including zero-day). The detection only works at the edge of attack originating cloud.} &  \parbox{2cm}{\cite{latanicki}\cite{youcant}\cite{botcloudGraham} \cite{botcloudBadis}\cite{mauro2}}\\\hline

\parbox{2cm}{Resource usage (D5)}  		& \parbox{3.5cm}{ OS level/hypervisor level detection methods to monitor abnormal usage}  & \parbox{3.5cm}{Interpreting the high utilization whether it is due to attack or due to the real traffic}  &  \parbox{4cm}{Only gives a signal  about  the possibility of attack and requires supplementary detection mechanisms } &  \parbox{2cm} {\cite{DARAC}\cite{defend}\cite{latanicki}  \cite{canwebeat}}\\\hline

\end{tabular}
}\end{center}
\caption{{DDoS Attack Detection Techniques in Cloud: D1 Pattern Detection}}
\label{D1}
\end{table*}

\subsection{Anomaly Detection (D1)}
\label{pattern}
{Anomalous patterns are usually identified from packet traces, established connections, web access logs or request headers. The specific pattern to identify in the log or the trace is decided by attack traces and other past historic behaviors. Web behavior has been modeled using a large number of characteristics and metrics working upon those characteristics. Mostly, authors have used web behavior of normal web traffic as a benchmark pattern. This normal web behavior is collected from the period when the attack is not present. On the other hand, few contributions prepare attack behavior profile and than filter-out the attack traffic by learning based detection. Feature selection, dataset preparation and testing or profiling against these learned rules are the three important set of operations, involved in these detection strategies. }

Now, we discuss few important strategies related to DDoS attack anomaly detection in cloud computing. Idziorek et al.~\cite{idziorekdetecting} worked on web access logs and argued that legitimate web access patterns follow ``Zipf" distribution and based on the web access pattern training, they could identify outliers, which do not follow this distribution in pattern~\cite{idziorekdetecting}. On the other hand, authors in~\cite{covariance}, used the baseline profiling of various IP and TCP flags which entails the network behavior model. Authors proposed the detection of flooding in the cloud using the training of normal and abnormal traffic and used the covariance matrix approach to detect the anomaly. Amongst other approaches, Shamsolmoali et al.~\cite{shamsolmoali2014statistical} proposed statistical filtering based attack detection. Proposed approach calculates divergence between normal traffic and attacker traffic on the basis of Jensen-Shannon Divergence~\cite{gomez2000analysis}. Initially, they have used the traditional TTL based differentiation among the legitimate users and spoofed attackers. After IP spoofing filtering, they have applied the Jensen-Shannon Divergence to identify the anomalies in the traffic to achieve around 97\% accuracy. There are few performance issues with TTL based approach. TTL based filtering is not useful unless we have a large database of actual TTL values of genuine requests using probing~\cite{templeton2003detecting}. This has not been addressed by the work in~\cite{shamsolmoali2014statistical}. In~\cite{cbf}, authors derived the web behavior using IP and TCP header fields. By this, they could calculate the confidence value in detection strategy. 

\noindent The major idea of this work was the claim that IP address and TTL values are related to multiple past contributions; therefore, the same can be extended to other fields in IP and TCP headers and a score for each incoming packet can be calculated. Jeyanthi et al.~\cite{jeyanthi} have proposed an approach, where they proposed to detect the DDoS attack on the basis of entropy. This is supported by ``Helinger'' distance which differentiated between the attack and genuine traffic distributions. Authors have used traffic rate, entropy and by predicting arrival rates of incoming traffic based on history. Authors in~\cite{edosarmor} have demonstrated an application specific way of differentiating web requests based on their behavior on an e-commerce site. This work has created two client profiles, one for good clients and another one for bad clients. Based on user walk-through on pages, purchases, searches these profiles are created and priority of customers is decided. Resource access pattern by clients is the main idea to detect attackers. In~\cite{ddosdefense}, authors created normal web profile, which include HTTP and XML header features. The number of elements, content length and depth have been used to create normal user profiles. Outliers are identified, which deviate from these profiles. Authors in~\cite{timespent} argued that an attacker would not spend any time on a page but would request them like a flood. They have gathered TSP behavior of users as well as of bots and identified that the attackers’ TSP is mostly negligible or even if it is not near zero, it is constant or periodic.

{The most important strength of these attack detection techniques lies in the machine learning of the past history of benign traffic or the attack traffic. With the advent of the paradigms such as big data analytics and software defined networks these detection methods have gained much important in quick attack detection and monitoring. A detailed survey of detection techniques is presented for traditional infrastructures in~\cite{Mirk}. These techniques are now becoming popular for cloud targeted attacks. }

{The major challenges for detection techniques lie in the behavior identification in terms of features and their training. The most important evaluation criteria for these methods lie in the false alerts (positive and negatives)  they generate during the testing of the incoming traffic. Other important challenge lies in stopping the IP spoofing which can defeat many of the detection strategies. }

\subsection{Source/Spoof Trace (D2)}
{Multiple trace back algorithms have been proposed in the literature, which identify and stop the spoof attack by tracing the source. Source traceback schemes are employed to stop/detect the identity spoofing techniques.} {These techniques are important as most of the detection/prevention methods model the user behavior or profile based on some identity which is mostly an IP address in case of web access. In the attack cases where IP spoofing is employed, the detection mechanisms can be defeated very easily. } 
\begin{table}[ht!]
\begin{center}
\centering
{
\begin{tabular}{|r|c|l|}\hline
\rotatebox{90}{\it \bf \parbox{1.5cm}{\small Solution category}} & \rotatebox{90}{\it \bf Contribution}&{\it \bf Major theme of the contribution}\\ \hline
\multirow{8}{*}{\rotatebox{90}{\parbox{5cm}{\small \bf Anomaly Detection (D1)} \hspace{-11mm}}} 
										& \small \cite{idziorekdetecting}  & \small Anomaly traffic detection using Zipf's law   \\ \cline{2-3}
										& \small \cite{covariance} & \small Co-variance profiling of IP/TCP flags\cite{gomez2000analysis} \\ \cline{2-3}
										& \small \cite{shamsolmoali2014statistical} & \small Filtering based on Jensen-Shannon Divergence   \\ \cline{2-3}
										&  \small \cite{xiao2015detecting}  & \small Co-relation based attack flow analysis   \\ \cline{2-3}
										& \small \cite{cbf} & \small IP/TCP flags based confidence filtering   \\ \cline{2-3}
										& \small \cite{jeyanthi} & \small  ``Helinger'' distance based multi-stage solution   \\ \cline{2-3}
										& \small \cite{edosarmor} & \small  User profiling using walk-through on site pages\\ \cline{2-3}
										& \small \cite{ddosdefense} & \small Filtering using SOAP headers  \\ \cline{2-3}
										& \small \cite{Idziorek11}  & \small  Identification of a genuine web session \\ \cline{2-3}
										& \small \cite{timespent} & \small  Profiling based on time spent on the pages\\ \hline	
\multirow{8}{*}{\rotatebox{90}{\parbox{4cm}{\small  \bf Source and Spoof Trace (D2)}}} 
										& \small \cite{chonka} & \small Back propagation neural networks tracing \\ \cline{2-3}
										& \small \cite{yang} & \small SOA-Based Trace back to reconstruct the path  \\ \cline{2-3}
										& \small \cite{osanaiye2015short} & \small OS fingerprinting to stop IP spoofing   \\ \cline{2-3}
										& \small \cite{multistage} & \small  Multi-stage source checking using text puzzles \\ \cline{2-3}
										& \small\cite{sourceauthentication} & \small Source authentication using token at each router\\ \cline{2-3}
										& \small \cite{luo2013preventing} & \small Source tracing based on location parameters \\ \cline{2-3}
										& \small \cite{yu2016feasible} & \small  Deterministic packet marking of ingress routers \\ \cline{2-3}
										&  \small \cite{Zhang}  & \small Multiple filters to stop spoofing   \\ \cline{2-3}		
										& \small \cite{templeton2003detecting}  & \small TTL probing to find genuine TTLs 	\\ \cline{2-3}							
										& \small \cite{law2005you} & \small Statistical filtering based spoof detection    \\ \hline			
\multirow{6}{*}{\rotatebox{90}{\parbox{2.5cm}{ \small \bf Count Based Filtering (D3)}}} 
										&  \small \cite{comber} & \small Hop count and request frequency thresholds  \\ \cline{2-3}
										&  \small \cite{enhanced} & \small TTL matching to detect IP spoofing \\  \cline{2-3}
										& \small \cite{index} &  \small Request threshold for a human in unit time \\  \cline{2-3}
										&  \small \cite{deflate} &  \small Threshold on number of connections by a source  \\  \cline{2-3}
										& \small \cite{templeton2003detecting}  & \small TTL probing to find genuine TTLs 	\\ \cline{2-3}
										&  \small \cite{army} & \small Request count threshold by each source  \\ \hline							
\multirow{5}{*}{\rotatebox{90}{\parbox{2cm}{\small \bf BotCloud Detection (D4)} }} 
										& \small \cite{latanicki} & Network/VMM checks to find attack VMs \small   \\ \cline{2-3}
										& \small \cite{youcant} & \small CSP driven attack flow check and source trace   \\ \cline{2-3}
										& \small \cite{botcloudGraham} & \small Bot detection in VMs using NetFlow   \\ \cline{2-3}
										& \small \cite{botcloudBadis} & \small Hypervisor led collaborative egress detection\\ \cline{2-3}
										& \small \cite{mauro2} & \small Virtual Machine Introspection (VMI) \\ \hline
\multirow{4}{*}{\rotatebox{90}{\parbox{2cm}{\small  \bf Resource Usage (D5)} \hspace{-2mm}}} 
										&  \small \cite{defend} & \small VM resource utilization threshold for detection \\  \cline{2-3}
										&   \small \cite{latanicki} & \small Resource counters and traffic thresholds for VMs \\  \cline{2-3}
										&  \small \cite{ATOM}  & \small  Resource usage anomalies and introspection\\  \cline{2-3}
										&   \small \cite{DARAC} & \small DDoS Aware auto-scaling to combat EDoS  \\ \cline{2-3}
										&  \small \cite{canwebeat} & \small Resource usage of attack target servers \\ \hline		
\end{tabular}
}
\end{center}
\vspace{-2mm}
\caption{{DDoS Attack Detection Techniques based on Pattern Detection}}
\label{D2}
\vspace{-6mm}
\end{table}

{Let us have a look at some techniques related to this subclass of solutions.} In~\cite{chonka} authors have done the same for SOAP requests.  Authors have used back propagation neural networks to tackle both the popular variants of the DDoS attack, which are HTML DoS and XML DoS. Authors in ~\cite{yang} drops all spoofed packets at edge routers using egress filtering. 

\noindent Authors proposed a method to identify the source of the attack by ``Service Oriented Architecture (SOA)" based technique. They proposed a source trace back method by introducing an additional server before real web server. This additional server is known as SBTA (SOA-Based Trace back Approach), which marks each packet by cloud trace back tag and also reconstructs the path to know the source. The proposed method uses a database to store and compare each incoming packet, and it requires an additional server to mitigate the attack. Osanaiye et al. in ~\cite{osanaiye2015short} have proposed an IP spoofing detection method, which is based on matching OS versions of both attackers and real IP owners. Authors have argued that the OS fingerprint of the spoofed attacker can be found out by asking the real OS fingerprint from the owner. Source authentication approaches have also been used in~\cite{sourceauthentication}, where a cryptographic token can be verified at each router to authenticate the source. Source checking approach has also been used in ~\cite{multistage}. Source traceback approaches are also dealt by hop count or TTL values which we discuss in Section~\ref{thresh}. Other important contributions include tracing sources by location~\cite{luo2013preventing} and statistical filtering~\cite{law2005you}. There are major surveys available in this direction where works related to botnets, their trends, and detection methods~\cite{botnetsurveysilva}.

{The source traceback and spoof identification methods are very important for all the detection methods. However, being a cooperative detection mechanism, these methods require a support from many other network devices such as edge routers, and services. Additionally, IP address being a ``source provided address'', it is extremely difficult to design spoof protection against massive spoofing by large scale botnets. }

\subsection{Count Based Filtering (D3)}
\label{thresh}
{This specific classification on ``Count Based Filtering" also fits in few attack prevention mechanisms as well, however, many a times thresholds are used to detect the initialization of attack and later can be used to identify the presence of the attack. The parameters on which these count thresholds are applied are basically network resources such as hop-count, number of connections or number of requests in a unit time from a single source.}

Authors in~\cite{comber} proposed the detection scheme, where apart from other schemes, a hop count filter has been used to identify spoofed packets. Similarly, authors in ~\cite{enhanced} have used TTL values alone for the purpose of DDoS prevention cum detection. As per this work, TTL values corresponding to various IP addresses are stored in white and black lists. If there is a new request then it is sent to graphical Turing test and on the basis of verification, it is added to the white list or black list. Those who are in white-list but with a different TTL, are also sent to the Turing test and on success their TTL value is updated. Authors in this paper extended their earlier work of the EDoS Shield~\cite{edosshield} and improved it for the case of IP spoofing. Their solution is based on hop-count diversity, where attacker packets are claimed to have same hop count, and thus they can be detected. In this strategy, if a user sends N request in period P, access to this user is only allowed, if his request count is less than threshold $T_{H}$. 

Authors in~\cite{index} used request count on the basis of human behavior and dropped all subsequent requests from the same IP for a finite period. Authors in~\cite{comber} have proposed a method to mitigate HTML and XML DDoS attacks by multiple level filtering on the basis of client puzzles, hop count and packet frequency. Various filters at server side incur significant overheads and latency for ordinary users. Similarly, authors in~\cite{army}, used the request count method to identify attackers and blacklist them. {DDoS Deflate~\cite{deflate} is a popular open source DDoS detection tool which is dependent upon the threshold of number of connections established by each source.} 

{The major strength of these solutions lie in their easy deployment and support by the available OS level firewalls such as \texttt{iptables} and APF. These methods also give administrators a quick control over the situation. }{However, these methods may not suite the requirements of all the users as the thresholds for a whole domain behind the NAT may not be similar to the thresholds required for dependent web-services. Additionally, methods such as TTL/hop-count requires a user database which has the actual hop-count/TTLs. Other issues arise due to a variety of heterogeneous implementations of hop-count in different systems. On the other hand, the IP spoofing techniques may defeat  the (non-TTL) schemes. Overall, the false positives or negative are important performance issues related to these count based filtering approaches.}

\subsection{BotCloud Detection (D4)}
\label{BotCloud}
{Any cloud DDoS attacker may also use cloud infrastructure for its own nefarious purpose. Cloud infrastructure can be used for the purpose of installing botnets. These clouds are known as BotClouds. This subcategory describes the contributions which tries to find or detect the internal attack VMs in the cloud network. Most of these BotCloud related solutions are source based or Cloud Service Provider (CSP) based approaches. }

{Authors in~\cite{latanicki} have presented a cloud level detection method to identify if there are attacker bots running inside hosted VMs. This has been achieved by network level and VMM level checks. Another contribution in this direction applies Virtual Machine Introspection (VMI) and data mining techniques to separate the infected VMs from other VMs in multi-tenant VMs~\cite{mauro2}. Authors had prepared a list of typical actions of malware bots infected VMs and used a clustering algorithm to identify the infected VMs based on the training. There are other BotCloud related solutions available in~\cite{youcant}~\cite{botcloudGraham}~\cite{botcloudBadis}. Authors in~\cite{youcant} provide a solution where the cloud provider checks the traffic flow and perform the anomaly detection using source traceback techniques at the network. Authors in\cite{botcloudGraham} provide a solution based on SDN approaches using Bot detection with the help of NetFlow protocol. Hypervisor based checks are used to detect the vulnerabilities in the guest VMs in~\cite{botcloudBadis} where collaborative egress detection technique is employed. Advanced methods such as one in~\cite{mauro2} propose a detection using virtual machine introspection (VMI). }

{The major strength of these methods lie in their deployment at the CSP end. By this, CSP has a control to monitor at the network edge for any anomaly in the traffic behavior or other performance counters. }

{However, these methods are not capable of detecting all kinds of attack flows such as zero-day or stealthy flows. On the other hand this kind of detection methods only work at the edge of attack originating cloud. In case, the CSP does not provide support for such detections, these attacks may become massive utilizing the profound resources of cloud computing. }

\subsection{Resource Usage (D5)}
{Utilization of various resource of the cloud or a physical server by a VM can also provide important information about the presence of the DDoS attack or an anticipation of the upcoming DDoS attack. Cloud environments run Infrastructure as a Service cloud using virtualized servers where hypervisor can monitor the resource usage of each VM on physical server. Once these VMs start reaching the decided resource utilization thresholds, the possibility of an attack can be suspected. }

In ~\cite{defend}, authors provided solutions on the basis of available resources with VMs and their upcoming requirements. Similarly,~\cite{latanicki} used performance counters and traffic to identify resource usage of VM and devise possible mitigation of the attack. Resource utilization possesses a very important and indirect metric to identify the possibility of an attack. Authors in~\cite{canwebeat} used resource limits as the sole method of the DDoS detection and then proposed mitigation methods.  {Authors in the~\cite{DARAC} implemented a DDoS aware resource allocation strategy in which the overloaded VMs are not directly flagged for resource increase. Instead, authors propose to segregate the traffic and increase the resources only on the basic of the demands of genuine flagged requests. Authors in~\cite{ATOM} have modeled the resource usage anomalies of VMs using virtual machine introspection to detect the possibility of resource surge due the DDoS attack. }

{DDoS attacks being resource intensive attacks provide a indirect relationship for the success of these resource usage based profiling and detection methods. Auto-scaling mechanisms are triggered on the basis of ``overload'' and ``underload'' states of the targeted VMs. This aspect also provide a possible co-relation between the VM resource usage and a DDoS originated resource surge. }

{The limitation of these set of approached lies in interpretation of the high resource utilization. It is very difficult to conclude whether the resource surge is due to the attack or due to the real traffic. As the resource surge only gives a alert about the possible resource surge, we may require other supplementary detection mechanisms. }

{After discussing the attack detection solution at length, it is clear that the traffic filtering based on the attack patterns is a major part of the DDoS attack solutions. Most of the methods are based on machine learning artifacts and provides a way to control the input traffic. However, the detection methods alone may not suffice for the purpose of integral protection from the DDoS attacks. The role of attack prevention solutions for the first hand protection and the role of attack mitigation solutions to ensure the resource availability for effective mitigation, can not be ignored. }

\begin{table*}[t]
\begin{center}
\centering
{
\begin{tabular}{|l|l|l|l|l|}\hline
{\it \bf  Techniques}	&	{\it \bf Strengths }		& 	{\it \bf Challenges} 		& {\it \bf Limitations} 			& {\it \bf Contributions}\\ \hline

\parbox{2cm}{Resource Scaling (M1)}  		& \parbox{3.5cm}{ Provides a quick relief to resource bottlenecks resource bottlenecks }  & \parbox{3.5cm}{Correctly deciding  whether and when extra resources are required}  &  \parbox{4cm}{False alarms may lead to EDoS. Co-hosted VMs may also be affected} &  \parbox{2cm} {\cite{RAA}\cite{alqahtaniddos}\cite{canwebeat} \cite{latanicki}\cite{DARAC}}\\\hline

\parbox{2cm}{Victim Migration (M2)}  		& \parbox{3.5cm}{Migrating the DDoS  victim service to other servers which helps in minimizing losses}  & \parbox{3.5cm}{Migration candidate selection and migration host selection }  &  \parbox{4cm}{Migration costs and overheads. Subsequent migrations/swaps in cloud} &  \parbox{2cm} {\cite{defend}\cite{latanicki}\cite{CDNonDemand}}\\\hline

\parbox{2cm}{OS Resource Management (ORM) (M3)}  		& \parbox{3.5cm}{Minimize the resource contention formed due to the attack at the victim service-end to have timely attack mitigation}  & \parbox{3.5cm}{Better checks needed to ensure the availability of contention}  &  \parbox{4cm}{Quick and dirty checks to ensure the availability of contention. It may affect the performance of the victim servers due to containment } &  \parbox{2cm} {\cite{Annals}\cite{VSC}}\\\hline

\parbox{2cm}{Software Defined Networking (SDN)(M4)	}  		& \parbox{3.5cm}{Abstract and timely view of the network and the incoming traffic using controllers}  & \parbox{3.5cm}{SDN may itself become an easy target of the DDoS attacks}  &  \parbox{4cm}{Mostly useful at network boundaries and ISP level network control } &  \parbox{2cm} {\cite{sahay2015towards}\cite{wang2015sdsnm}\cite{cloudddosSDN} \cite{Cloudddossdn3}\cite{yan2015distributed}\cite{wang2015sdsnm}}\\\hline

\parbox{2cm}{DDoS Mitigation as a Service (DMaaS)(M5)}  		& \parbox{3.5cm}{Cloud based hybrid mitigation using extra resources or remote traffic monitoring and prevention services}  & \parbox{3.5cm}{Cost overhead issues. Methods are mostly similar to the on-premise solutions but mitigation expertise is an advantage}  &  \parbox{4cm}{Solutions may not cater various kinds of applications and attacks. Local issues may not be visualized by DDoS mitigation-as-a-service} &  \parbox{2cm} {\cite{DaaS1}\cite{DaaS2}\cite{guenane2014reducing} \cite{amazondiscussionforum}\cite{Cloudwatch}	}\\\hline

\end{tabular}
}\end{center}
\vspace{-2mm}
\caption{{DDoS Attack Mitigation (M) Techniques in Cloud}}
\label{M1}
\vspace{-6mm}
\end{table*}

\begin{table}[t]
\begin{center}
\centering
{
\begin{tabular}{|r|c|l|}\hline
\rotatebox{90}{\it \bf \parbox{2cm}{Solution \\category}} & \rotatebox{90}{\it \bf Contribution}&{\it \bf Major theme of the contribution}\\ \hline
\multirow{5}{*}{\rotatebox{90}{\parbox{2.5cm}{ \small \bf Resource Scaling (M1)}}} 
										& \small \cite{alqahtaniddos} & \small Multi-level (VM, service, application and cloud)  \\  \cline{2-3}
										& \small \cite{canwebeat}  & \small Dynamic resource scaling for quick detection  \\  \cline{2-3}
										& \small \cite{latanicki} &  \small Resource scaling in federated clouds   \\  \cline{2-3}
										&  \small  \cite{awsddosprotection}  & \small Scaling to absorb the attack \\ \cline{2-3}
										& \small \cite{DARAC} & \small Scaling based on capacity planning \\ \cline{2-3}
										& \small \cite{CDNonDemand}  & \small Scaling over low cost untrusted CDN clouds \\ \hline						
\multirow{4}{*}{\rotatebox{90}{\parbox{2cm}{\small  \bf Migration (M2)} \hspace{-5mm}}} 
										& \small \cite{defend} & \small Victim VM migration to other physical servers \\ \cline{2-3}
										& \small \cite{MigrationFuji} & \small  Migrating proxy entry points at overlay   \\  \cline{2-3}
										& \small \cite{latanicki} & \small Victim VM migration to other physical servers  \\  \cline{2-3}
										& \small \cite{Wang2014} & Exploiting VM migrations using DDoS \\ \hline						
\multirow{2}{*}{\rotatebox{90}{ \parbox{1.5cm}{\small  \bf ORM (M3)} \hspace{-8mm}}}
										& \small \cite{Annals} & \small Service resizing to reduce the resource contention \\ \cline{2-3}
										& \small \cite{VSC} & \small Containment to reduce the resource contention   \\ \hline
\multirow{6}{*}{\rotatebox{90}{\parbox{1.5cm}{\small  \bf SDN (M4)} \hspace{-7mm}}}
										& \small \cite{sahay2015towards} & \small  ISP-level monitoring of traffic and routing\\ \cline{2-3}
										& \small \cite{wang2015sdsnm} & \small Strict authentication and access control  \\  \cline{2-3}
										& \small \cite{Cloudddossdn3} &  \small Re-configurable network monitoring and control\\  \cline{2-3}
										& \small \cite{tsai2017defending} &  \small SDN based deep packet inspection \\ \hline		
\multirow{9}{*}{\rotatebox{90}{\parbox{2cm}{\small  \bf DMaaS (M5)} \hspace{-15mm}}} 
										& \small \cite{DaaS1} & Victim cloud-based network service \small   \\  \cline{2-3}
										& \small \cite{DaaS2} & \small Proof-of-work scheme and ephemeral servers \\ \cline{2-3}
										& \small \cite{guenane2014reducing}  & \small Hybrid (On-premise firewall plus Cloud firewall) \\  \cline{2-3}
										& \small \cite{amazondiscussionforum} & \small Resource caps to limit the attack effects \\  \cline{2-3}
										& \small \cite{Cloudwatch} & \small Cloud metric monitoring and alarms   \\ \hline																
\end{tabular}
}
\end{center}
\vspace{-2mm}
\caption{{DDoS Attack Mitigation Techniques in Cloud}}
\label{M2}
\vspace{-5mm}
\end{table}

\section{Attack Mitigation (M)}
\label{mitigation}
In this section, we have grouped all methods which would help a victim server to continue serving requests in the presence of an attack. Downtime is a major business parameter for websites and an organization may lose a significant number of prospective customers~\cite{economiclosses}. In this section, we have grouped methods, which would allow victim server to keep serving requests in the presence of an attack.  Mitigation and recovery are complementary to each other to keep the server alive, which is under the attack. These methods are used temporarily and once the attack subside, the server may be brought back to the actual situation. 

\noindent Most of mitigation and recovery methods, which are proposed here are purely related to infrastructure clouds and their solutions are in the direction of mitigating EDoS attacks.  {We further classify this section into five subcategories:
\begin{enumerate}[1.]
\item Resource Scaling.
\item Victim Migration.
\item OS Resource Management (ORM).
\item Software Defined Networking (SDN).
\item DDoS Mitigation as a Service (DMaaS).
\end{enumerate} 
For a quick view, the overall theme of each set of the classified methods, their strengths, challenges, and weaknesses are listed in Table~\ref{M1}. We also prepare a list of important individual contributions in Table~\ref{M2} where we enlist a brief theme of each solution to show the variety of contributions available in each subclass.}
 
\subsection{Resource Scaling (M1)}

{Dynamic auto-scaling of resources is one of the most popular features of the clouds. It is also treated as one of best mitigation methods to counter DDoS attack allowing server availability or continuity with scaled resources. Auto scaling can be done horizontally, where new instances may be started on the same or different physical server to serve incoming requests till the victim server is facing the attack. In vertical scaling, resources like CPU, memory and disk can be scaled in the same VM or the same logical unit. These extra resources can help the victim machine to survive the attack and keep running. One of the major disadvantages of this strategy is that it can become an advantage for the attacker to increase the attack strength to even deplete added resources and generating a requirement of more resources shaping the attack into an EDoS~\cite{RAA}. }

We now discuss few important contributions related to attack mitigation and recovery using resource scaling. Authors in~\cite{alqahtaniddos} proposed a multi-level DDoS detection system for web services. VM owner level (Tenant level), service Level, application level and cloud level detection are placed to have a collaborative DDoS detection system. It is one of those solutions which are utilizing the information from all the stakeholders in mitigating the DDoS attacks. However, there might be large overhead and other security concerns due to information flow among multiple levels. 

One of the first and most important contributions in this area, which touches cloud-specific issues is by Shui Yu et al.~\cite{canwebeat}. Authors in this paper considered the dynamic resource allocation feature of the cloud to help the victim server to get additional resources for DDoS mitigation. In this way, individual cloud customers are saved from DDoS attacks by dynamic resource allocation. Experiments on real website data sets show that their queuing theory based scheme work to mitigate DDoS attack. Authors in~\cite{latanicki} presented three different scenarios to stop the DDoS attack in the cloud. These three scenarios include external attacks to internal servers, internal attacks to internal servers and internal attacks to external servers. Authors provided strategies to detect the attack and get recovered using scaling and migrations in a federated cloud environment. Reserved resources are kept in~\cite{canwebeat} to support the server in attack times. ``How much reserved resources should be kept?'' is an important question. The cost of additional and idle resources is a drawback. It is one of the flexibility which keeps back up and reserved resources for a rainy day~\cite{latanicki}. On the other hand, authors in~\cite{DARAC} have provided a resource allocation strategy which do not scale the resources on DDoS generated resource surges. 
\newpage
\noindent Authors in~\cite{CDNonDemand} have provided a mechanism which uses low-cost untrusted cloud servers in the presence of DDoS attacks to scale services frugally. ``CDN On Demand'' is an open source platform developed to support the mechanism~\cite{CDNonDemand}. {Industry solutions such as~\cite{awsddosprotection} also advocate for quick resource scaling for quick attack absorption. }

{The resource scaling is an important aspect of cloud computing which is also useful in quick attack mitigation while maintaining the service availability. The resource scaling is a process which is useful for attacker to recover by expanding the VM resources or VM instances. }

{However, the resource scaling may also become against the overall idea of cost-savings using the cloud hosting. In case the attacks are stealthy and remains undetected, than the resource scaling may increase the attack costs multi-fold. A detailed discussion on the role of resource scaling and the mitigation costs in~\cite{IEEECC}. }

\subsection{Victim Migration (M2)}
{VM migration has changed the way the entire running server is shifted to another physical server without noticeable downtime. Migration can be used to shift the victim server to a different physical server, which is isolated from the attack and once the DDoS is detected and mitigated, the server can again be shifted back to the actual place. }

We discuss few important contributions using victim migration to mitigate the DDoS attack. Authors in~\cite{defend} proposed a similar strategy by keeping some reserved resources on a server. While the attack is detected, they migrate the victim to those reserved resources and bring it back when attack ceases. Downtime to legitimate customers is one issue which is very important while migration is chosen as a mitigation method. Additionally, if the attack continues for longer duration or repeated, the cost considerations will be high. Authors in~\cite{latanicki} also used a similar approach. Authors in~\cite{defend} have proposed a remedial method for the server affected by DDoS to keep it in the running or serving state. DDoS attack has been detected at the level of Virtual Machine Monitor (VMM) instead of any count based or packet filtering. VMM is detecting the possibility of the DDoS attack by continuously monitoring resource utilization levels. Once the resource utilization levels reach a certain threshold, VMM flags a DDoS attack. On signaling, VMM migrates or duplicates the running VM as well the application to a separate isolated environment on the same physical server. This isolated environment is created with the help of reserving some additional resources for backup, where the ``victim'' is shifted in case of the DDoS attack. Once the attack gets over, the isolated environment again shifts the VM back to its real place. {On the other hand, there are characterizations which shows the exploitation of VM migrations using DDoS attacks~\cite{Wang2014}. Authors in~\cite{somani2016ddos} have shown in their characterization that the DDoS attacks may lead to migrations which may even spread the collateral damages from one physical server to other physical servers. Other solutions such as~\cite{MigrationFuji} show a different flavor migration using migrating proxy entry points at overlay networks at the victim server-end. }

{Victim migration to backup resource provides a way to control the attack effects and employ the attack mitigation. Also it may help in scaling the services using migrating to a large sized candidate/host servers where the migratee server can use the additional resources to detect and mitigate the attack.}

{There are few issues related to the sustainability of these schemes. In particular, wastage of additional resources, which has to be available all the time is a major issue. Detection of the DDoS just by keeping a watch over resource utilization might not be a good idea, as there might be higher utilization because of real traffic during flash events or heavy computation needs. In fact, this behavior might lead to an unnecessary duplication to an isolated environment. Even the overhead of duplicating the system when the attack is evident might not be a wise step to overcome the security of the server and application. The contribution in~\cite{defend} has overlooked one very important aspect about DDoS attack which is attack duration. If the attack continues, how would server serve its legitimate consumers who are trying to access the service at that point in time? If it does not serve them then ``for how long, the service will be down?'' is an important factor. Additionally, if it serves them then there is a large overhead of transferring states and keeping data and sessions up to date.}

\subsection{{OS Level Resource Management  (M3)} }
\label{ORM}
{There are few recent contributions in the DDoS attack solution space for cloud computing which deals with resource management at the level of VM operating systems. These OS level resource management methods argues that DDoS attacks being the resource intensive attacks may affect the overall mitigation methods running inside the victim VMs. By minimizing the contention at the level of the operating systems, the mitigation and recovery can be expedited. }

{Authors in~\cite{Annals} show a service resizing based methods where once the attack is detected, the victim service is affined to the minimum processing units (CPUs) using OS level controls. Authors have shown that a DDoS attack may become an ``extreme DDoS'' attack if the resource contention becomes severe. This contention may even delay the overall mitigation process. Author extend this service resizing using victim resource containment in~\cite{VSC} where using OS control groups are used to contain or isolate the victim service. Authors have also shown collateral effects on other critical service co-hosted with the victim service on the same operating system. }

{These local resource management methods are shown to minimize the resource contention formed due to the attack at the victim service-end to have timely attack mitigation. Authors have shown important metrics related to attack mitigation in terms of attack detection time, mitigation time and the reporting time with some additional features such as attack cooling down period which they optimize using TCP tuning. }

{The major limitation of these approaches lies in their quick and dirty checks to ensure the availability of resource contention. These methods may also affect the performance of the victim servers due to the resource containment with an additional cost of the resources. }

\subsection{Software Defined Networking (M4)}
\label{SDN}
{Software Defined Networking (SDN) is an emerging reconfigurable network paradigm which may change the whole DDoS mitigation space. SDN in its core separates data and control planes of switching to support the network reconfigurability on the fly. }

There are few initial and ongoing works related to SDN assisted DDoS mitigation mechanisms. Authors in~\cite{sahay2015towards} have proposed a SDN-based solution in which ISP-level monitoring of traffic and routing of malicious traffic is done to specially designed secure switches. In this work, the victim is required to request ISP for DDoS mitigation. ISP having an abstract view of incoming traffic applies the traffic labeling using OpenFlow switches. The suspicious traffic is then redirected to security middle-boxes which apply access policies on the traffic. Authors have left the detection and mitigation part on the customer side. {A similar proposal by authors in~\cite{wang2015sdsnm}, suggested a prototype implementation of SDN-based detection mechanism. The major idea of this work lies in the strict access control policies for the incoming traffic which requires strict authentication for each incoming request. Advanced deep packet inspection based approaches using SDN are discussed in~\cite{tsai2017defending}. A detailed tutorial and guideline of SDN-based solutions are given in~\cite{cloudddosSDN}.  } 

{SDN as a paradigm has immense possibilities of support for the attack mitigation for massive as well low-rate DDoS attacks due to its reconfigurability and quick networks view and monitoring. }

{Mitigation Solutions utilizing SDN capabilities are still evolving and may become very helpful due to their important features. However, studies such as~\cite{yan2015distributed} show that even the SDN infrastructure itself can become a victim of DDoS attacks. }

\subsection{{DDoS Mitigation as a Service (DMaaS) (M5)}}
{There are multiple cloud based service/third party services which are are capable of providing the DDoS protection~\cite{prolexic}~\cite{Akamai}~\cite{arbor}. Mostly, DDoS protection is done on a server or an intermediate node forwarding packets to the server. There are solutions which are hosted in the cloud and provide DDoS mitigation as a service~\cite{DaaS1}~\cite{DaaS2}. Multiple providers in the market offer this facility. However, all these mitigation methods are threshold/count based or human intervention based. }

On the other hand, there are not many specific products available to mitigate DDoS targeting a cloud. Authors in~\cite{guenane2014reducing} proposed a DDoS mitigation service. This service is intended to help the physical on-premise firewall to do the mitigation quickly. The proposed solution is termed as a hybrid firewall, which uses both physical firewall and virtual firewall (placed in the cloud). Amazon has started providing resource limits on EC2 instances to provide an initial solution. There were multiple requests from consumers to cloud providers about keeping cap or limit on maximum allowed resources and subsequently there were additions from cloud providers related to resource consumption limit alerts to customers~\cite{amazondiscussionforum}. Additionally, Amazon has created a service, cloudWatch~\cite{Cloudwatch}, to provide real-time information about various metrics towards a service hosted in Amazon cloud so that necessary steps can be taken up.  

{Third party mitigation services or DDoS mitigation as a Service may become very helpful for attack mitigation and recovery using a on-premise tools and/or cloud based solution. The attack mitigation history and expertise in handling various attacks may become helpful for enterprises seeking specialized help. Also the cloud based service may also utilize the extensive resource support available in the cloud. }

{The major limitation of these DMaaS approaches include remote mitigation which may not fasten the mitigation process. Additionally, victim service owners may not want to share the control with the third parties due to the privacy issues of their traffic and the business logic. Other important aspects include the cost of the solutions and the sustainability requirements of the victim enterprises. In addition to all the above five categories of mitigation methods, shutdown is a typical trivial method to stop the DDoS attack on a server. But this method does not provide any solution to downtime of the service which is non-negotiable. In some approaches, the victim server is started at another place as a new instance and present instance is shut down. This helps in starting a synced clone at another place. Though there are high chances that the attacker will also attack the new server. A similar idea has been proposed in~\cite{moving} where attacked proxy servers are shutdown and the traffic is redirected to new proxy servers. }

{Attack mitigation methods narrated above provide a detailed overview of various attack mitigation and recovery solutions available in cloud computing space. The mitigation methods are usually a supportive layer of protection for the attack prevention and detection solutions. \\As discussed above, for the case of cloud computing, the mitigation methods play very important role due to their applicability to resource management during the attack. }

\section{Discussion and Future Directions}
\label{discussion}
{There is a large volume of work which has been referred while preparing this survey. With this rigorous survey, it is clear that most of the works, which have emerged in this domain are concentrating on the following five aspects:
\begin{enumerate}
\item Characterization or Impact study.
\item Prevention using Turing Tests.
\item Threshold or pattern based filtering.
\item Support to stop IP spoofing.
\item Resource scaling.
\end{enumerate}}
Most of the solutions proposed so far are using one or a combination of the above approaches. There are only a few solutions which are including the auto scaling, multi-tenancy and utility model into account. {The cloud computing infrastructure may be used to build effective mitigation solutions which ensure the quick attack mitigation and timely recovery to ensure effective service availability.}

\subsection{Solution Considerations}
In order to offer an effective solution to DDoS in the cloud, the following features require special treatment. Here, each feature has been discussed with an intention to provide an aid to the ideal solution. 
\subsubsection{Auto-scaling}
Auto-scaling in the cloud is usually triggered by monitored metrics of a VM or an application running inside a VM. These are resource usage metrics like CPU, memory and bandwidth and other counters like response time, query processing time etc. Triggering the auto-scaling would either result in an increase or decrease in allocated resources. Controlling Auto-scaling or false triggering of auto-scaling requires specific checks which can verify the real usage. These checks can be conducted at VM level, hypervisor level or even at abstract cloud level~\cite{DARAC}. 
\begin{itemize}
\item Vertical Scaling: This feature deals with the scaling on a physical server where multiple VMs are running with co-hosted isolations. Vertical scaling would deal with adding or removing resources on these VMs. Total resources which are available on the physical server are fixed but each VM may have a different amount of resources at different times. This really depends on the resource allocation policy and the SLA. Any DDoS affected VM would continuously request for more and more resources and available idle resources (with the Cloud Service Provider) should fulfill these requests. This decision is critical as it would also decide the health of co-hosted VMs and cost considerations of newly added resources~\cite{somani2016ddos}.
\item Horizontal Scaling: This scheme allows adding new instances of the same VM at other physical servers. These instances are created to share the load and maintain the quality of web services. An ideal composite scaling strategy would first rely on vertical scaling followed by horizontal scaling. The decision-making process to start more instances on more servers should look for a true need and cost considerations. Another important point in horizontal scaling is limiting the maximum number of instances of an application. This can be decided by the cloud consumer but a restriction on it may lead to losing business.
\end{itemize}
\subsubsection{Multi-tenancy}
Multi-tenancy leads to proper hardware utilization of high-capacity servers which would have been underutilized if not implemented as multi-tenant environments. Vertical scaling would have much flexibility in case few VMs are running on a single machine. On the other hand, cloud providers would have ROI (Return on Investment) considerations and would want to host more and more VMs. Other than this, performance isolation and performance interference aspects should also be looked carefully while designing capacity of these servers. DDoS defense mechanism and its design should reflect protecting multi-tenant environments. 
\subsubsection{Pay-as-you-go model}
Pay-as-you-go model is advantageous for both consumers and providers. Literature has mostly counted pay-as-you-go models as an advantage for consumers. But this becomes advantageous for a cloud provider when VMs it has hosted in its cloud requires more and more resources on a regular basis. In case, this additional requirement is fulfilled than the consumer needs to pay for additional resources and provider gets benefited. Almost all solutions should keep the accounting and billing model in the perspective while designing cost-aware DDoS defense solutions. 
\subsubsection{Migration}
As described in the Section~\ref{mitigation}, VM migration is a very important method to minimize effects of the DDoS in a virtualized cloud. Migrations incur a cost in terms of downtime, configuration changes, and bandwidth usage. If the application does not have the capability to start more instances to share the load, migration is the only way to minimize the downtime and denial of service. As horizontal scaling cannot be done in such cases, the duration for which DDoS attacks lasts would also play a major role. Large attack duration may lead to multiple subsequent migrations here and there~\cite{somani2016ddos}, and thus a large number of side-effects to the cloud and other VMs. DDoS defense mechanism should be able to minimize the number of migrations during the attack period by closely working with horizontal scaling.
\subsubsection{Solution Costs}
{The most important motivation for the enterprises to shift their service to cloud infrastructure is the cost effectiveness. However, we have seen in the detailed attack effects (Figure~\ref{effects}), that the DDoS attack losses may become multi-fold in the cloud infrastructure as compare to traditional on-premise infrastructure. The major portion of the cloud users include small and medium enterprises which necessitates the sustainability or budget factor as important aspect while designing the solutions. Authors in~\cite{IEEECC} have detailed the cost considerations for DDoS attack solutions.}

\begin{figure*}[htb]
\begin{center}
\includegraphics [width=0.85\textwidth]{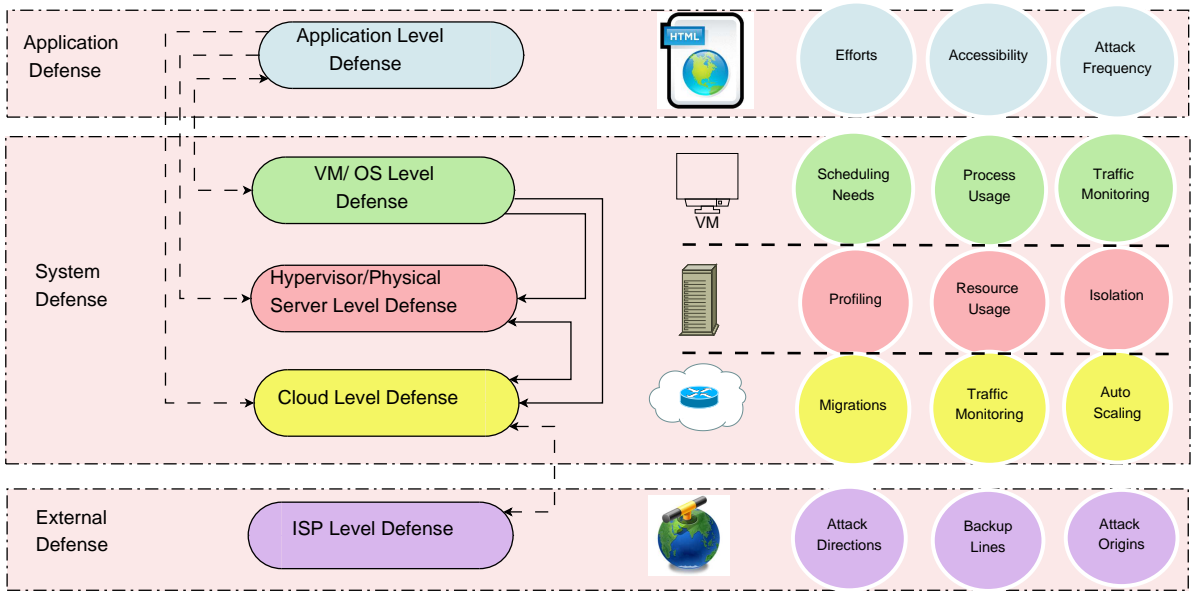}
\vspace{-2mm}
\caption{Solution Hierarchy with Three Solution Levels}
\label{Effectivesolution}
\end{center}
\vspace{-8mm}
\end{figure*}
\subsection{Building an effective solution}
\label{effective}
In this section, we are compiling details related to effective solutions towards DDoS in the cloud. Even though these solutions only outline the solution space and related issues but they also give a systematic design of an ideal solution. The model shown in Figure~\ref{Effectivesolution} illustrates levels of solutions, where the defense mechanism deployed.  {We show five defense levels in this figure where we also show the important services/information or monitored metrics provided at each level. The information flow among these five levels is a tricky part because of the security questions related to the business logic and access control. This is a design question and can be solved by allowing only anonymized monitoring data to be transferred among these five levels. In Figure~\ref{Effectivesolution}, we have also pointed out specific solution design features and aspects, which can be dealt at each level.} 
\subsubsection{Application level defense}
Applications are the front ends where attackers send requests. These applications are mostly web services which send web pages on the basis of user's HTTP requests. TCP SYN and ICMP floods are also sent to applications responding to them. The defense mechanism should lookout for an unexpected increase in a number of requests from a set of source IP addresses. Identifying these source IP addresses is the root solution to the DDoS detection problem. Most of the solutions available in the literature try to defend at application level\cite{spow}\cite{edosshield}. These solutions include Turing tests, request frequency and hop count based filtering. For efficiency, the following three design aspects are considered. 
\begin{itemize}
\item Mitigation Efforts: {\it The effort required to identify and prevent an attacker is usually more than the effort required to serve the attacker as a normal user.} Many times the complex defense schemes checks for multiple filters and each request has to go through these filters. A lengthy and complex mechanism may incur large computation and storage costs. Defense mechanism may get into an ``Indirect EDoS" due to heavy filtering efforts and result into puzzle accumulation attacks~\cite{toolsdown}. 
\item Accessibility: Accessibility of a normal user should not be compromised. Many times usability of a web service is affected because of special mechanisms of tests and other defense mechanisms. Usability aspects are very important and surveys have shown that even a delay of 1 second in page loading time may affect conversion rate~\cite{conversion}. This may lead to higher response times and usability aspects for many users and especially for elderly or differently abled persons. 
\item Request rates: Attacks may not always be high rate obvious flooding attacks. Low rate but continuous attacks may also affect a server's economic aspects~\cite{Idziorek11}. In~\cite{Idziorek11}, authors have shown that sending one request per minute for a month also incurs a cost.
\end{itemize}
{Other important metrics and solution requirements for application level defense is already discussed in Section~\ref{mitigation}. Most of the solutions related to attack detection (Section~\ref{detection}) and attack mitigation (Section~\ref{mitigation}) solutions are applicable to the VM/OS level, Hypervisor level and the Cloud level defense. }
\subsubsection{VM/OS level defense}
VM running on a hypervisor runs a complete operating system on top of them. An eye over the resource usage of specific processes, their generation and object fetch cycles may provide a clue about the attack monitoring. Many consequences of the EDoS occur due to the decisions taken at this level. At present, there are very few solutions in this direction~\cite{alqahtaniddos}. In addition to all these features, performance isolation is one of the most important assurance which is required at this level~\cite{somani2016ddos}. {Local resource management at the level of the victim operating systems can be very effective in managing the DDoS attacks~\cite{Annals}\cite{VSC}. These solutions advocates of minimizing the resource contention created due to the DDoS attacks which may help in minimizing the overall downtime (detailed in Section~\ref{ORM}). }
\subsubsection{Hypervisor level defense}
A hypervisor is the control and management layer (a bare metal hypervisor like XenServer) which handles the most important task of ``Vertical Scaling". Scheduling VMs, managing their memory and storage are some of the most important areas where an effective monitoring mechanism could be employed. Additionally, this level can be controlled by the ``Cloud" level which can send/receive important alerts and take appropriate decisions.
There are some mitigation solutions which have partially used this level for defense~\cite{alqahtaniddos}\cite{latanicki}. 
\subsubsection{Cloud level defense}
Cloud level defense may want to look at the amount of traffic coming-in and going-out to have a top level abstract idea of the attack. At this level, any anomaly in the normal behavior can be detected. Additionally, decisions regarding ``Horizontal Scaling" are also taken at this level which take migrations and cost into considerations. A solution which involves communication between hypervisor and cloud manager would be a good design to deal with the defense mechanisms. Few novel solutions, which are based on cloud level of defense are~\cite{jia2014catch}\cite{canwebeat}\cite{alqahtaniddos}\cite{latanicki}\cite{guenane2014reducing}. {Network level defense capabilities provided by SDN infrastructures can also become very helpful at this level to gain the quick control of the network (detailed in Section~\ref{SDN}).} 
\subsubsection{ISP level defense}
Defense at the ISP level~\cite{ISP} can be of immediate help for DDoS attacks originating from specific networks. Projects like Digital Attack Map~\cite{map} may be used as a handy reference here. Even a choked line by a DDoS can be replaced by an ISP by another backup line for recovery from the attack. Both cloud and ISP level should keep a close watch over the incoming/outgoing traffic generated to limit them by some mechanisms. DDoS target networks, as well as DDoS originating networks, may be identified by the ISP collaborations. Authors in~\cite{chen2005perimeter} have proposed a method which works on filtering at edge routers of the network. Authors have proposed three mechanisms to provide ISP supported DDoS defense mechanisms. Authors have shown the effectiveness of the mechanism through simulations. Authors argued that the perimeter based mitigation method can sustain the defense against attacks even if 40\% of the customer networks are ``attacker'' networks. Another important work in this direction is proposed by~\cite{chen2007collaborative}. The major idea of this work is to look for abrupt traffic changes across networks using attack-transit routers at ISP networks. These changes are modeled and detected using distributed change-point detection mechanism utilizing special constructs known as change aggregation trees (CAT). Authors have also given a trust policy among networks to cure these attacks collaboratively. 
{
In Figure~\ref{Effectivesolution}, we show three defense abstractions which are formed using the five defense levels discussed above. 
\begin{enumerate}
\item Application Defense which is formed using attack prevention mechanisms.  
\item System Defense which is formed by three defense levels (VM/OS, hypervisor and cloud).
\item External Defense which is formed using ISP level and third party defense. 
\end{enumerate}
Considering above facts, we provide following major solution designs to DDoS attacks in cloud computing.
\begin{enumerate}
\item Application Defense : This is a design which considers defense at application level only. This level of defense is the most used and helps in the multi-tenant environment where each hosted VM should be isolated because of multiple virtualization and data security threats. Most solutions in the literature follow this design but it is also clear that this design alone is not suitable to take care of aspects of cloud.
\item Application Defense +  System Defense : If this design can be implemented with ease than it can be proven as one of the effective solutions as the defense mechanism would take advantage of the information from multiple sub-levels in ``System Defense''. The information gathered and supplied by the Level ``Application Defense'' would be important in taking pro-active decisions at level ``System Defense''.
\item System Defense/System Defense + External Defense : Both of these solutions would work at the system level to defend the DDoS attack. The difference is that the later will use the ISP support. ``System Defense''  alone would be effective, however, identifying the ``True positives" and ``False Negatives" is the most important concern here. As without actual verification of attack traffic from ``Application Defense' level, this defense would lack effectiveness.
\item Application Defense + System Defense + External Defense : This is a complete design with multi-level support and information or alert flow. After solving the data security and business logic theft issues among levels, it would be an ideal design solution for an Infrastructure cloud.
\end{enumerate}}
{We have shown various performance and evaluation metrics related to the DDoS attack solutions in Table~\ref{effects}. All those metrics are applicable here and may be used for creating effective solutions as per design abstractions discussed above. }
\section{Summary and Conclusions}
\label{conc}
{This work provides a comprehensive and detailed survey about the DDoS attacks and defense mechanisms eventually available in the cloud computing environment. We have shown through the discussion that EDoS attack is a primary form of DDoS attack in the cloud. DDoS attacks have important characteristics which play an important role while considering utility computing models. This paper introduces the cloud computing features which are critical in order to understand the DDoS attack and its impact. }

{We have also presented attack statistics, its impact, and characterization by various contributors. We propose a novel comprehensive taxonomy of DDoS attack defense solutions in cloud computing. We believe that this survey would help to provide a directional guidance towards requirements of DDoS defense mechanisms and a guideline towards a unified and effective solution. There are a large number of solutions which have targeted the DDoS attack from one of the three solution categories of attack prevention, detection, and mitigation. Among these solutions, there are few contributions which are targeting at cloud-specific features like resource allocation, on-demand resources, botcloud detection, and network reconfiguration using SDNs. We also provide a comprehensive list of performance metrics of these solution classes for their evaluation and comparison. We believe that this novel attempt of presenting the complete set of evaluation metrics for a variety of DDoS solutions may help in orchestrating the benchmarking of upcoming solutions.}

{At the end, we have provided a detailed guideline for effective solution design. This effective solution guideline provides a complete view of solution design space and parameters to help future defense mechanisms. This survey may play an important role in providing the basis for the innovative and effective solutions to prevent and deter DDoS attacks in cloud computing. Characterization at the level of a cloud as a whole and multiple clouds would really help in understanding the impact of this attack at a larger level. As discussed in the survey, multi-level solutions specifically designed for cloud and its features would surely perform better as compared to traditional DDoS solutions. Cost and attack aware resource allocation algorithms in the cloud would help in mitigating the attack. Finally, the multi-layer solution guideline based solutions can be tested to have their effective evaluation in cloud infrastructure.}
\vspace{10mm}\\
{\bf \noindent Acknowledgments}\\
Gaurav Somani is supported by a Teacher Fellowship under Faculty Development Program funded by University Grants Commission under XII Plan (2012-2017). This work is also supported by SAFAL (Security Analysis Framework for Android Platform) project funded by Department of Electronics and Information Technology, Government of India. Mauro Conti is supported by a Marie Curie Fellowship funded by the European Commission (agreement PCIG11-GA-2012-321980). This work is also partially supported by the EU TagItSmart! Project (agreement H2020-ICT30-2015-688061), the EU-India REACH Project (agreement ICI+/2014/342-896), the Italian MIUR-PRIN TENACE Project (agreement 20103P34XC), and by the Project “Tackling Mobile Malware with Innovative Machine Learning Techniques” funded by the University of Padua. Rajkumar Buyya is supported by Melbourne-Chindia Cloud Computing (MC3) Research Network and the Australian Research Council via Future Fellowship program.
\vspace{-4mm}
\bibliographystyle{elsarticle-num}
\bibliography{referencethesis} 
\end{document}